\documentclass[aps,prb,showpacs,twocolumn,amsmath,amssymb,superscriptaddress,letterpaper]{revtex4-1}

\usepackage[dvips]{graphicx}
\usepackage{latexsym}
\usepackage{amsmath}
\usepackage{amsfonts}
\usepackage{amssymb}
\usepackage{bm}
\usepackage{color}
\usepackage{txfonts}
\usepackage{multirow}
\usepackage{array}
\usepackage{threeparttable}
\usepackage{makecell}

\begin{document}
	\newcommand{\fig}[2]{\includegraphics[width=#1]{#2}}
	\newcommand{\la}{{\langle}}
	\newcommand{\ra}{{\rangle}}
	\newcommand{\dg}{{\dagger}}
	\newcommand{\upa}{{\uparrow}}
	\newcommand{\dna}{{\downarrow}}
	\newcommand{\ab}{{\alpha\beta}}
	\newcommand{\ias}{{i\alpha\sigma}}
	\newcommand{\ibs}{{i\beta\sigma}}
	\newcommand{\hH}{\hat{H}}
	\newcommand{\hn}{\hat{n}}
	\newcommand{\hc}{{\hat{\chi}}}
	\newcommand{\hU}{{\hat{U}}}
	\newcommand{\hV}{{\hat{V}}}
	\newcommand{\br}{{\bf r}}
	\newcommand{\bk}{{{\bf k}}}
	\newcommand{\bq}{{{\bf q}}}
	\def\gsim{~\rlap{$>$}{\lower 1.0ex\hbox{$\sim$}}}
	\setlength{\unitlength}{1mm}
	\newcommand{{\vhf}}{\chi^\text{v}_f}
	\newcommand{{\vhd}}{\chi^\text{v}_d}
	\newcommand{{\vpd}}{\Delta^\text{v}_d}
	\newcommand{{\ved}}{\epsilon^\text{v}_d}
	\newcommand{{\vved}}{\varepsilon^\text{v}_d}
	\newcommand{{\tr}}{{\rm tr}}
	\newcommand{\pprl}{Phys. Rev. Lett. \ }
	\newcommand{\pprb}{Phys. Rev. {B}}

\title{Topological Superconductivity in an s-wave Superconductor and Its Implication to Iron-based Superconductors }

\author{Shengshan Qin}
\affiliation{Kavli Institute of Theoretical Sciences, University of Chinese Academy of Sciences, Beijing, 100049, China}
\affiliation{CAS Center for Excellence in Topological Quantum Computation, University of Chinese Academy of Sciences, Beijing, 100049, China}

\author{Chen Fang}\email{cfang@iphy.ac.cn}
\affiliation{Beijing National Research Center for Condensed Matter Physics,
and Institute of Physics, Chinese Academy of Sciences, Beijing 100190, China}
\affiliation{Kavli Institute of Theoretical Sciences, University of Chinese Academy of Sciences, Beijing, 100049, China}
\affiliation{CAS Center for Excellence in Topological Quantum Computation, University of Chinese Academy of Sciences, Beijing, 100049, China}

\author{Fu-Chun Zhang}
\affiliation{Kavli Institute of Theoretical Sciences, University of Chinese Academy of Sciences, Beijing, 100049, China}
\affiliation{CAS Center for Excellence in Topological Quantum Computation, University of Chinese Academy of Sciences, Beijing, 100049, China}

\author{Jiangping Hu}\email{jphu@iphy.ac.cn}
\affiliation{Beijing National Research Center for Condensed Matter Physics,
and Institute of Physics, Chinese Academy of Sciences, Beijing 100190, China}
\affiliation{Kavli Institute of Theoretical Sciences, University of Chinese Academy of Sciences, Beijing, 100049, China}
\affiliation{CAS Center for Excellence in Topological Quantum Computation, University of Chinese Academy of Sciences, Beijing, 100049, China}

\date{\today}

\begin{abstract}
In the presence of both space and time reversal symmetries, an s-wave $A_{1g}$ superconducting state  is usually topologically trivial. Here we demonstrate that an exception can take place in a type of nonsymmorphic  lattice structures.  We specify the demonstration in  a system with a centrosymmetric space group $P4/nmm$, the symmetry that governs iron-based superconductors, by showing the existence of a second-order topological state protected by a mirror symmetry. The topological superconductivity is featured by $2 \mathcal{Z}$ degenerate Dirac cones on the $(10)$ edge, and $\mathcal{Z}$ pairs of Majorana modes at the intersection between the $(11)$ and $(1 \overline{1})$ edges.  The topological invariance and Fermi surface criterion for  the topological state are provided.  Moreover,  we point out that the previously proposed s-wave state in iron-based superconductors, which features a sign-changed superconducting order parameter between two electron pockets, is such a topological state. Thus, these results not only open a new route to pursue  topological superconductivity, but also establish a measurable quantity to  settle one long-lasting debate on the pairing nature of iron-based superconductors.
\end{abstract}
\maketitle


\textit{Introduction.}
Stimulated by the potential application in fault-tolerant quantum computation, the search for topological superconductors\cite{RevModPhys.80.1083, RevModPhys.82.3045, RevModPhys.83.1057, RevModPhys.88.035005, Kitaev_2001, Alicea_2012, hao2019topological, wu2020pursuit} has been one central topic in condensed matter physics. After decades' efforts, great progress has been made both theoretically\cite{PhysRevLett.100.096407, PhysRevLett.104.040502, PhysRevLett.105.077001, PhysRevLett.105.177002, PhysRevLett.105.046803, PhysRevB.82.115120, PhysRevLett.107.097001, PhysRevLett.111.047006, PhysRevLett.111.056402, PhysRevB.88.155420, PhysRevLett.115.127003, li2016manipulating, PhysRevLett.117.047001, yang2016majorana, PhysRevX.9.011033} and experimentally\cite{TSC_Ando, das2012zero, nadj2014observation, MZM_Jia, wang2018evidence, vortex_Feng, vortex_Kong, machida2019zero, chen2019observation, kong2020tunable, wang2020evidence, MZM_flux}. In recent years, motivated by the study in topological insulators\cite{doi:10.1146/annurev-conmatphys-031214-014501, PhysRevLett.106.106802, hsieh2012topological, tanaka2012experimental, bradlyn2017topological, po2017symmetry, PhysRevX.7.041069, song2018quantitative, PhysRevX.8.031070, zhang2019catalogue, tang2019efficient} more and more attention has been being paid to the topological superconducting phases protected by the crystalline symmetries, dubbed as the topological crystalline superconductors. The crystalline symmetries play a dramatic role in classifying the topological superconductors. On one hand, the crystalline symmetries greatly enrich the classification. For example, in the presence of mirror symmetry\cite{PhysRevLett.111.087002, doi:10.7566/JPSJ.82.113707, PhysRevLett.111.056403, PhysRevLett.112.106401}, rotational symmetry\cite{rotation_Sato, rotation_CFang, rotation_RXZhang, PhysRevLett.123.027003, PhysRevLett.122.207001, qin2019topological, rotation_HOTDS} or glide mirror symmetry\cite{PhysRevB.93.020505, PhysRevB.93.195413, PhysRevLett.122.227001}, many new topological superconducting states beyond the Altland-Zirnbauer classification\cite{classification1, classification2} have been identified. On the other hand, the crystalline symmetries have strict constraints on the values of the topological indices. For instance, if a class-D\uppercase\expandafter{\romannumeral3} superconductor is centrosymmetric with the superconducting order belonging to a trivial irreducible representation, it can hardly carry any topological nontrivial property. Thus,  normally, abundant s-wave centrosymmetric superconductors  can not be  topological superconductors.


In this letter, we show that the s-wave superconducting states  in centrosymmetric superconductors can carry nontrivial topology in the presence of additional nonsymmorphic symmetries, $i.e.$ the glide mirror symmetry or the screw rotation symmetry. This exception stems from anomalous band degeneracies induced by the nonsymmorphic symmetries. We specify the study with the space group $P4/nmm$ ($\#.129$), the nonsymmorphic symmetry group that governs iron-based superconductors. A second-order topological superconducting state  in the $A_{1g}$ pairing channel is constructed and is characterized by a $2 \mathcal{Z}$ winding number protected by the mirror symmetry.  The state hosts $2 \mathcal{Z}$ degenerate Dirac cones on the edge and $\mathcal{Z}$ pairs of Majorana modes at the corner, where the corresponding mirror symmetry is preserved. We develop a Fermi surface criterion for such topological superconductors.  The  previously proposed s-wave state for iron-based superconductors\cite{PhysRevB.84.024529, Hirschfeld_2011}, which are characterized by a sign-changed superconducting order parameter between two electron pockets, belongs to this exceptional class of topological states.  Thus, the theory also establishes a directly measurable quantity to  reveal  the pairing nature in iron-based superconductors.

\begin{figure}[!htbp]
	\centering
	\includegraphics[width=\linewidth]{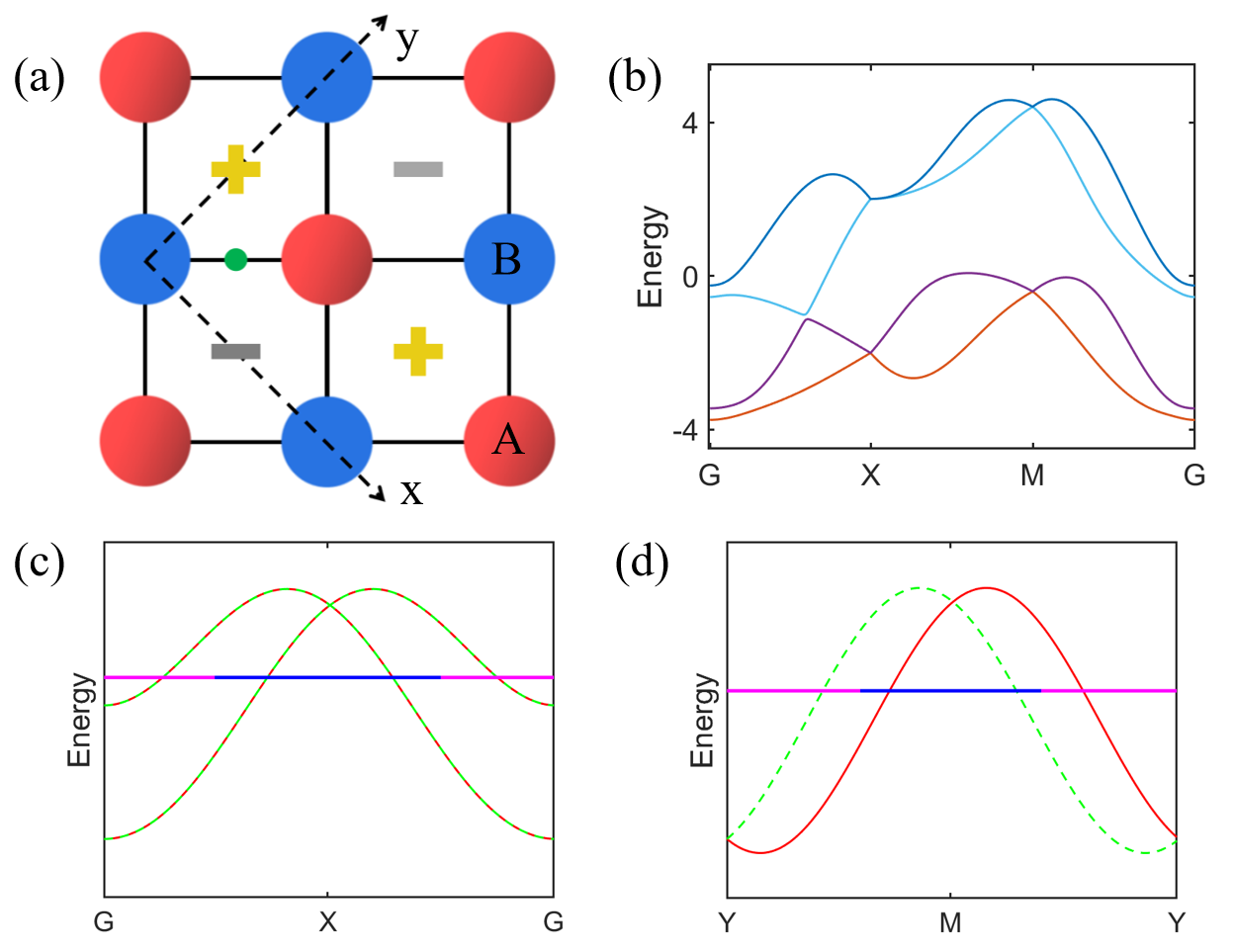}
	\caption{\label{fig1} (color online) (a) shows the monolayer case for the lattice structure respecting the space group $G = P4/nmm$. The translational symmetry from sublattice A to B is broken by an intrinsic effective electric field normal to the $xy$ plane allowed by the symmetry\cite{zhang2014hidden, wu2017direct, Zhang_2020}, with $+/-$ labeling the direction.  (b) shows the band structures corresponding to $\mathcal{H}_0$ in Eq.\eqref{normal_Hamiltonian}, with G, X, M, Y the $(0, 0)$, $(\pi, 0)$, $(\pi, \pi)$, $(0, \pi)$ points in the first Brillouin zone. (c) and (d) sketch the bands in different mirror $\{ M_y | {\bf 0} \}$ invariant subspaces (represented by  red lines and green dashed lines) along the $\Sigma_{\text Y}$ and $\Sigma_{\text G}$ respectively. Pink and blue represent different pairing signs in a s-wave superconducting state to illustrate the second-order topological superconductivity protected by the mirror symmetry $\{ M_y | {\bf 0} \}$. Notice that only the condition in (d) corresponds to the topological case. }
\end{figure}

\textit{Group structure of $P4/nmm$.}
We start with a brief review of the structure of the space group $G = P4/nmm$.  We consider a quasi-2D lattice structure shown in Fig.\ref{fig1}(a), similar to the monolayer iron-based superconductors. To describe the symmetry group, we adopt the \textit{Seitz operators}, $\{ \alpha | {\bf \tau} \}$, which acts on the lattice in the way $\{ \alpha | {\bf \tau} \} {\bf r} = \alpha{\bf r} + {\bf \tau}$. It is easy to check the Seitz operators have the following properties
\begin{eqnarray}\label{Seitz}
\{ \alpha | {\bf \tau} \}^{-1} &=& \{ \alpha^{-1} | -\alpha^{-1} {\bf \tau} \}, \nonumber\\
\{ \alpha_1 | {\bf \tau_1} \} \{ \alpha_2 | {\bf \tau_2} \} &=& \{ \alpha_1 \alpha_2 | \alpha_1 {\bf \tau_2} + {\bf \tau_1} \},
\end{eqnarray}
where $\alpha$ is a point group operation and $\tau$ is a spatial translation. Apparently, the translation symmetry in the space group can be denoted as $\{ E | t_1{\bf a_1}+t_2{\bf a_2} \}$, with  the primitive lattice translations ${\bf a_i}$  and integers $t_i$.  The quotient group $G/T$, with $T$ being the translation group, is specified by 16 symmetry operations and is expressed instructively as\cite{PhysRevX.3.031004}
\begin{align}\label{quotientG}
\begin{split}
G/T &= D_{2d} \otimes Z_2,
\end{split}
\end{align}
with $D_{2d}$ the point group defined on the lattice sites, and $Z_2$ a two-element group including the inversion symmetry $I$ defined at the center of the bond between two nearest lattice sites, as illustrated in Fig.\ref{fig1}(a). In Eq.\eqref{quotientG}, the quotient group is a direct product of the two subgroups in a sense that symmetry operations are equivalent if they differ by a lattice translation. Moreover, since the two subgroups are defined on two inequivalent points, $G$ is nonsymmorphic and there are glide mirror and screw rotation operations in $G/T$.

\textit{Symmetry constraints on Bloch states.}
First, we analyze the constrain of the electronic structures  by the  nonsymmorphic symmetries  on the boundaries of the Brillouin zone. We compare  the $\Sigma_{\text Y}$ (M-Y-M) line  with  the $\Sigma_{\text G}$ (X-G-X) line in the Brillouin zone.   the $\Sigma_{ {\text Y} / \text{G}}$ line is invariant under the group $C_{2v} \otimes Z_2$, which includes the following symmetry operations
\begin{eqnarray}\label{MYM}
&& \{ E | {\bf 0} \}, \ \ \quad \{ C_{2z} | {\bf 0} \}, \quad \{ I | {\bf \tau_0} \}, \qquad \{ M_z | {\bf \tau_0} \}, \nonumber \\
&& \{ M_x | {\bf 0} \}, \quad \{ M_y | {\bf 0} \}, \quad \{ C_{2x} | {\bf \tau_0} \}, \quad \{ C_{2y} | {\bf \tau_0} \},
\end{eqnarray}
where ${\bf \tau_0} = {\bf a_1}/2 + {\bf a_2}/2$, and all the symmetry operations are defined at the lattice site in Fig.\ref{fig1}(a).  Besides the crystalline symmetries, we can also consider the operation  $\{ I | {\bf \tau_0} \} \Theta$ with $\Theta$ being the time reversal symmetry. $\{ I | {\bf \tau_0} \} \Theta$ is antiunitary and leaves every ${\bf k}$ point unchanged in the Brillouin zone. We can use the above symmetries to classify the eigenstates on $\Sigma_{ {\text Y} / \text{G}}$. Here, we focus on the mirror symmetry $\{ M_y | {\bf 0} \}$, which allows us to specify eigenvalues of Bloch states as  $m_y = \pm i$. In the spinful condition, the symmetries satisfy
\begin{eqnarray}\label{MYM_commutation1}
(\{ I | {\bf \tau_0} \} \Theta) (\{ I | {\bf \tau_0} \} \Theta) | \varphi({\bf k}) \rangle = - | \varphi({\bf k}) \rangle,
\qquad \quad \nonumber \\
(\{ I | {\bf \tau_0} \} \Theta) \{ M_y | {\bf 0} \} | \varphi({\bf k}) \rangle = e^{ik_y} \{ M_y | {\bf 0} \} (\{ I | {\bf \tau_0} \} \Theta) | \varphi({\bf k}) \rangle,
\end{eqnarray}
namely
\begin{align}\label{MYM_commutation2}
k_y = \pi: \quad & [ \{ I | {\bf \tau_0} \} \Theta, \{ M_y | {\bf 0} \} ]_+ = 0, \nonumber \\
k_y = 0: \quad & [ \{ I | {\bf \tau_0} \} \Theta, \{ M_y | {\bf 0} \} ]_- = 0,
\end{align}
where $[]_-$ and $[]_+$ label the commutation and anticommutation operations respectively\cite{footnote1}.  Based on Eqs.\eqref{MYM_commutation1}\eqref{MYM_commutation2},  for a Bloch state with eigenvalue $m_y$,   there is a  degenerate  state $-m_y$ on $\Sigma_{\text{G}}$ and a degenerate  state $m_y$ on $\Sigma_{\text{Y}}$.  Together with  the constraint of the time reversal symmetry which  maps a state with the mirror eigenvalue $m_y$ at ${\bf k}$ to a state with $-m_y$ at $-{\bf k}$, we  get a overall picture on the bands along the $\Sigma_{ \text{Y} / \text{G} }$ line as illustrated in Figs.\ref{fig1}(c)(d).

\textit{s-wave ($A_{1g}$) topological superconductivity.}
We consider  the topological classification of s-wave superconducting states protected by the mirror symmetry  $\{ M_y | {\bf 0} \}$. In class-D\uppercase\expandafter{\romannumeral3} superconductors\cite{classification1, classification2}, there always exists the chiral symmetry $\mathcal{C}$, which is the product of the time reversal symmetry and the particle-hole symmetry.  In the s-wave state, the chiral symmetry commutates with the crystalline symmetries and is preserved  in each of the $\{ M_y | {\bf 0} \}$ invariant subspaces.

Now we consider the two lines, $\Sigma_{\text{Y}}$ and $\Sigma_{\text{G}}$, in the Brillouin Zone, and treat them as two one dimensional (1D) subsystems. The winding number in each of the mirror subspaces along the $\Sigma_{ \text{Y} / \text{G} }$ line is given by
\begin{align}\label{winding}
\begin{split}
w_{ \text{Y} / \text{G} }^{\pm i} &= \int_{ \text{Y} / \text{G} } \frac{dk_x}{2\pi} tr[ \mathcal{C} \mathcal{H}_{\pm i}^{-1}({\bf k}) \partial_{k_x} \mathcal{H}_{\pm i}({\bf k}) ]
\end{split}
\end{align}
where $\mathcal{H}_{\pm i}(\bf k)$ is the superconducting Hamiltonian in the $m_y = \pm i$ mirror subspace. The winding numbers along the two lines are dramatically different due to the different symmetry-enforced band degeneracies along the two lines.  Considering  the time reversal and space inversion symmetries together with the constraints in Eq.\eqref{MYM_commutation2},  we can obtain the following conclusions (details in Supplementary Materials (SM)): (i) on $\Sigma_{\text{G}}$ in each mirror invariant subspace, the winding number is always trivial, namely $w_{\text{G}}^{+i} = -w_{\text{G}}^{-i} = 0$;  (ii) on $\Sigma_{\text{Y}}$ the classification is $w_{\text{Y}}^{+i} = -w_{\text{Y}}^{-i} = 2 \mathcal{Z}$. The results can be further understood by calculating the 1D winding numbers through the Fermi surface criterion\cite{PhysRevB.83.224511, PhysRevX.10.041014}
\begin{align}\label{FS_criterion}
\begin{split}
w &= \frac{1}{2} \sum_j  sgn[ v(k_F^j) \Delta(k_F^j) ],
\end{split}
\end{align}
where $v(k_F^j)$ and $\Delta(k_F^j)$ are the Fermi velocity and superconducting pairing at the $j$-th Fermi point, and $sgn[]$ is the sign function. For an s-wave centrosymmetric superconducting state, the pairing order is an even function of ${\bf k}$. The Fermi velocity on $\Sigma_{\text{G}}$ is odd of ${\bf k}$ in each mirror subspace  as indicated in Fig.\ref{fig1}(c), leading to $w_{\text{G}}^{\pm i} = 0$.  However,  on $\Sigma_{\text{Y}}$, the bands are no longer symmetric between ${\bf k}$ and $-{\bf k}$.  The nonzero $w_{\text{Y}}^{\pm i}$  can be generated  as indicated in Fig.\ref{fig1}(d).  Specifically, the topological number is characterized by $| w_{\text{Y}}^{\pm i} |$ Dirac cones degenerate at $k_y = \pi$ on the $(10)$ edge. We emphasize that the above topological superconductivity is unique for centrosymmetric superconductors governed by nonsymmorphic groups, since the anomalous band degeneracy in Fig.\ref{fig1}(d) can occur only in the presence of the nonsymmorphic symmetries.

\textit{Lattice model.}
To verify the above analysis, we construct a simple two-orbital ($p_x$ and $p_y$) model. The band structure with respect to the $P4/nmm$ symmetry can be generally described by
\begin{eqnarray}\label{normal_Hamiltonian}
\mathcal{H}_0 &=& t \cos k_x s_0 ( \sigma_0 + \sigma_3 ) \eta_0 + t \cos k_y s_0 ( \sigma_0 - \sigma_3 ) \eta_0 \nonumber \\
&+& 4 t_1 \cos\frac{k_x}{2} \cos\frac{k_y}{2} s_0\sigma_0\eta_1 - 4 t_2 \sin\frac{k_x}{2} \sin\frac{k_y}{2} s_0\sigma_1\eta_1 \nonumber \\
&-& 2 \lambda_R \sin k_x s_2 ( \sigma_0 + \sigma_3 ) \eta_3 - 2 \lambda_R \sin k_y s_1 ( \sigma_0 - \sigma_3 ) \eta_3 \nonumber \\
&+& \frac{\lambda}{2} s_3\sigma_2\eta_0,
\end{eqnarray}
where $s_i$, $\sigma_i$ and $\eta_i$ are the Pauli matrices standing for the spin, orbital and sublattice degrees of freedom respectively. In the model, $t$ is the nearest-neighbour (NN) intra-sublattice intraorbital hopping, $t_{1/2}$ is the NN inter-sublattice intraorbital/interorbital hopping, $\lambda$ is the atomic spin-orbit coupling and $\lambda_R$ is the Rashba-type spin-orbit coupling allowed by the symmetry group $P4/nmm$. It takes the form $i \lambda_R ({\bf d} \times {\bf s}) \cdot {\bf e_z}$ in the real space, with ${\bf d}$ the intra-sublattice NN vector and ${\bf e_z}$ the direction of the effective electric field shown in Fig.\ref{fig1}(a). Notice that we only preserve the intraorbital $\lambda_R$, with the $\pi$-bond type omitted. We emphasize that the effective electric field inducing $\lambda_R$ is an intrinsic result of the symmetry group. Such inversion-symmetric polarization has been identified both theoretically and experimentally in the previous studies\cite{zhang2014hidden, wu2017direct, Zhang_2020}. In iron-based superconductors, this term stems from the non-coplanar cations and anions.

We set the parameters in Eq.\eqref{normal_Hamiltonian} as $\{ t, t_1, t_2, \lambda, \lambda_R \} = \{ -1.0, 0.4, 0.6, 0.3, 0.75 \}$, and get the band structures in Fig.\ref{fig1}(b). A s-wave ($A_{1g}$ ) superconductivity is described generally by $\mathcal{H}_{\text{sc}} = [ \Delta_0 + 2\Delta_1 ( \cos k_x + \cos k_y ) ] s_0 \sigma_0 \eta_0$ in the basis $\psi^\dagger({\bf k}) = ( c^\dagger({\bf k}), is_2 c(-{\bf k}) )$, where $\Delta_0$ is the on-site intraorbital pairing and $\Delta_1$ is the NN intra-sublattice intraorbital pairing.

\begin{figure}[!htbp]
	\centering
	\includegraphics[width=\linewidth]{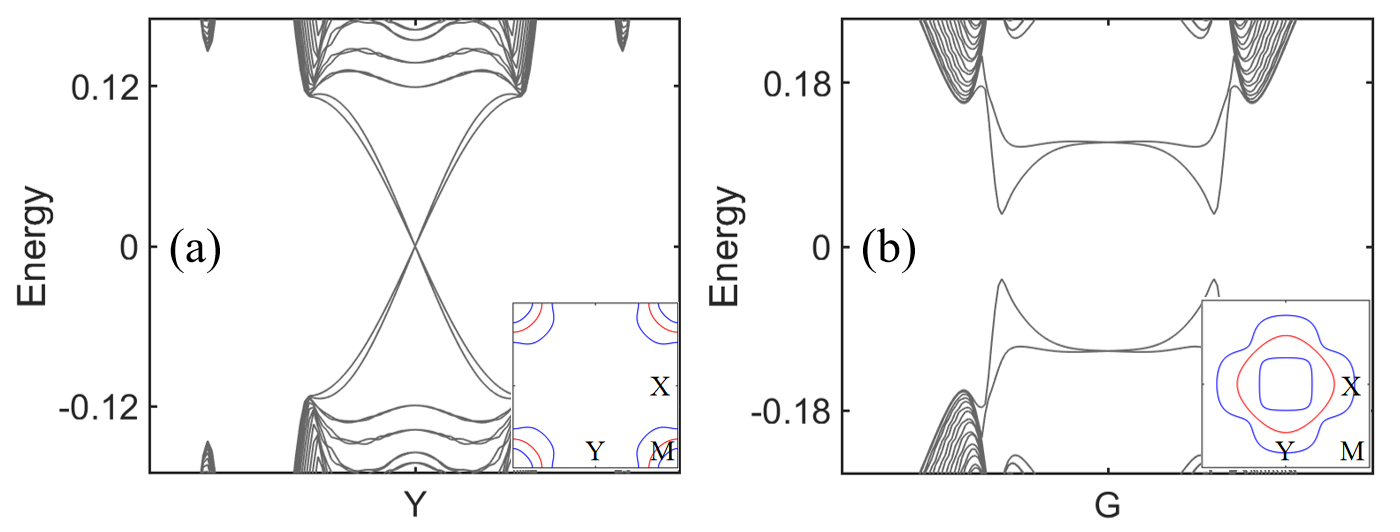}
	\caption{\label{fig2} (color online) Edge modes on the $(10)$ edge for the Hamiltonian in Eq.\eqref{normal_Hamiltonian} in the $A_{1g}$ pairing channel. The insets in (a)(b) are the Fermi surfaces (blue) in the normal bands, and the pairing nodes (red). The corresponding parameters are chosen as $\Delta_0 = 0.58$, $\Delta_1 = 0.2$ and the chemical potential $\mu = 3.6$ for (a), and $\Delta_0 = -0.3$, $\Delta_1 = 0.2$ and $\mu = 0.8$ for (b).
	}
\end{figure}

To  show the topological superconductivity protected by the mirror symmetry $\{ M_y | {\bf 0} \}$, we simulate the edge states on the boundary normal to mirror plane, $i.e.$ the $(10)$ edge.   There are two different conditions in which the Fermi surfaces and pairing nodes are located around (i) the $M$ point and (ii) the $G$ point in the Brillouin zone respectively,  as shown in Fig.\ref{fig2}.  In the first case there are two Dirac cones degenerate at $k_y = \pi$ on the boundary, shown in Fig.\ref{fig2}(a), indicating $| w_{\text{Y}}^{\pm i} | = 2$. In the second case, there is no gapless mode as   shown in Fig.\ref{fig2}(b), indicating a topological trivial state.  To see that the results are consistent with the Fermi surface criterion in Eq.\eqref{FS_criterion}, we draw the Fermi surfaces and the vanishing lines of the  superconducting orders, shown as the inserts in Figs.\ref{fig2}(a)(b).

\begin{figure}[!htbp]
	\centering
	\includegraphics[width=\linewidth]{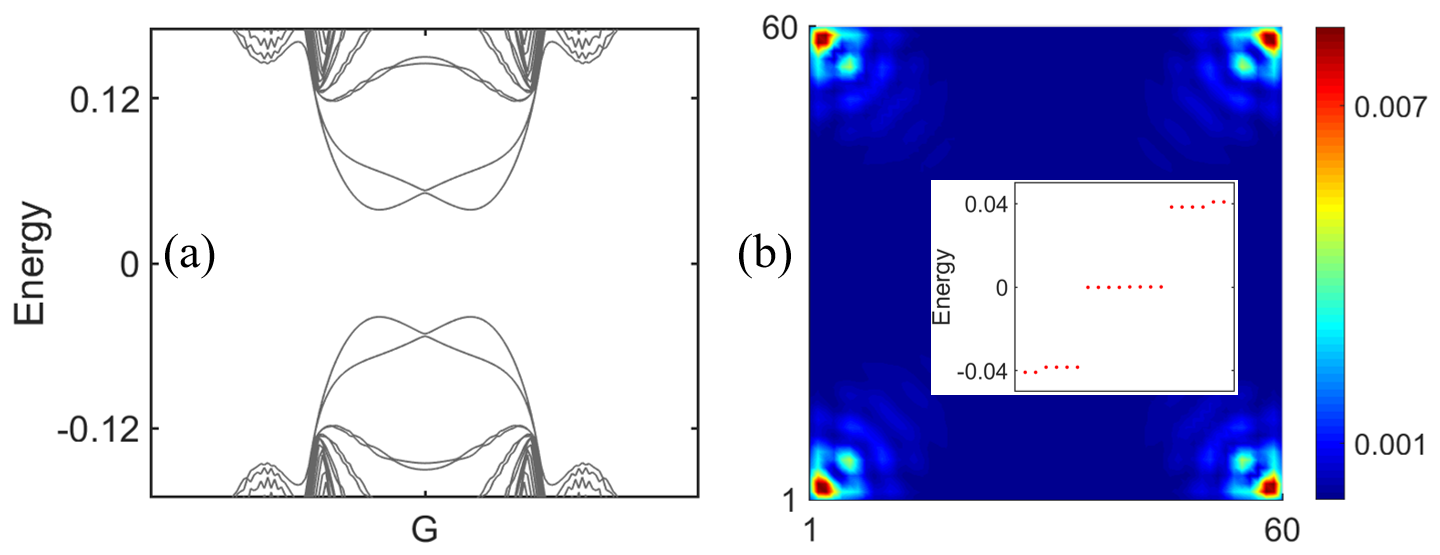}
	\caption{\label{fig3} (color online) (a) shows the edge modes on the $(11)$ boundary in the same condition with that in Fig.\ref{fig2}(a).  (b) shows the wavefunction profiles of the zero-energy modes in the real space under  open-boundary conditions both in the $(11)$ and $(1 \overline{1} )$ directions, with the low-energy spectrum presented in the inset.
	}
\end{figure}

\textit{Majorana corner modes.}
In fact, the mirror-protected topological superconductivity is a second-order topological state\cite{benalcazar2017quantized, PhysRevB.96.245115, PhysRevLett.119.246402, schindler2018higher, PhysRevLett.119.246401}. It supports Majorana modes at the corner with a $\mathcal{Z}$ classification equal to $| w_{\text{Y}}^{\pm i} | / 2$. To verify this, we first simulate the edge states on the $(11)$ edge under the same Fermi surface condition in Fig.\ref{fig2}(a).  The $(11)$ edge has no  $\{ M_y | {\bf 0} \}$ symmetry.  As expected, as shown in Fig.\ref{fig3}(a), no gapless modes survive because $\{ M_y | {\bf 0} \}$ is no longer maintained. Furthermore, we simulate the corner modes localized at the intersection between the $(11)$ and $(1 \overline{1} )$ edges. As shown in Fig.\ref{fig3}(b), at each corner there exists one pair of zero-energy modes, indicating second-order topological superconductivity.

The arise of the corner Majorana modes is guaranteed by the mirror symmetry $\{ M_y | {\bf 0} \}$. This can be clearly verified by considering the effective theory on the edges. We start with the gapless modes on the $(10)$ edge shown in Fig.\ref{fig2}(a), corresponding to $| w_{\text{Y}}^{\pm i} | = 2$. On this edge, there are the time reversal symmetry, the particle-hole symmetry, the chiral symmetry, and the mirror symmetry. With the constraints of these symmetries and a proper gauge choice, the effective theory can be written as $\mathcal{H}_{eff} = v k_y s_3 \kappa_3$, with $v$ the Fermi velocity, $s_i$ the Pauli matrices labeling the spin degree of freedom and $\kappa_i$ the Pauli matrices for the remaining degree for the edge Dirac cones (details in SM).

Now we gradually bend edge $(10)$ into a right angle, with the two sides along the $(11)$ and $(1 \overline{1} )$ directions. In this procedure, the mirror symmetry $\{ M_y | {\bf 0} \}$ is preserved, mapping the $(11)$ edge to the $(1 \overline{1} )$ edge. However, the gapless modes on each edge gain a mass and are gapped out, as the mirror symmetry is not maintained on the $(11)$/$(1 \overline{1} )$ edge. The gapped modes on these two edges take the form (details in SM)
\begin{eqnarray}\label{edge}
\mathcal{H}_{eff}^{(11)} &=& v k s_3 \kappa_3 + m_{ (11) } s_1 \kappa_3, \nonumber \\
\mathcal{H}_{eff}^{(1 \overline{1} )} &=& v k s_3 \kappa_3 + m_{ (1 \overline{1} ) } s_1 \kappa_3.
\end{eqnarray}
with $m_{ (11) / (1 \overline{1} ) }$ the mass term on the $(11)$/$(1 \overline{1} )$ edge. Moreover, the mirror symmetry $\{ M_y | {\bf 0} \}$ demands $m_{ (11) } = -m_{ (1 \overline{1} ) }$ (details in SM). Obviously, the theory in Eq.\eqref{edge} describes a massive Dirac theory, whose mass changes sign at the intersection between the $(11)$ and $(1 \overline{1} )$ edges. This mass domain leads to a pair of Majorana modes localized at the corner\cite{PhysRevD.13.3398, PhysRevLett.121.096803, PhysRevLett.121.186801, PhysRevLett.122.187001}, consistent with the results in Fig.\ref{fig3}(b).  Based on the analysis, it is clear that besides the intersection between the $(11)$ and $(1 \overline{1} )$ edges, any corner respecting the mirror symmetry $\{ M_y | {\bf 0} \}$ would support the Majorana modes. Moreover, under a general condition with $| w_{\text{Y}}^{\pm i} | = 2 \mathcal{Z}$, we can conclude that  $\mathcal{Z}$ Majorana Kramers' pairs would arise at the corner.

The above analysis  in 2D lattices can be generalized to the three dimensional (3D)  case. Specifically, we can stack the lattice in Fig.\ref{fig1}(a) along the $z$ direction, in which condition the space group $P4/nmm$ is preserved, and consider the edge modes on the $(100)$ surface and the hinge modes at the intersection between the $(110)$ and $(1 \overline{1} 0)$ surfaces. We can calculate the mirror-protected winding numbers on lines $(k_x, 0, k_z^0)$ and $(k_x, \pi, k_z^0)$ for each fixed $k_z^0$. Notice that the winding number is well defined on these lines because the chiral symmetry leaves each ${\bf k}$ point unchanged in the Brillouin zone. The mirror symmetry $\{ M_x | {\bf 0} \}$ requires similar normal-band degeneracies with that in Fig.\ref{fig1}(c)/(d) on line $(k_x, 0/\pi, k_z^0)$. Therefore, for the s-wave pairing state the edge and corner modes for each $k_z^0$ are expected to be similar to those in the 2D case. With all $k_z^0$ taken into consideration, it can be straightforwardly concluded that in the 3D topological nontrivial state, there must exist $2 \mathcal{Z}$ degenerate quasi-1D Dirac cones on the $(100)$ surface and $\mathcal{Z}$ pair of flat Majorana hinge modes at the corner between the $(110)$ and $(1 \overline{1} 0)$ surfaces.


\begin{figure}[!htbp]
	\centering
	\includegraphics[width=0.5\linewidth]{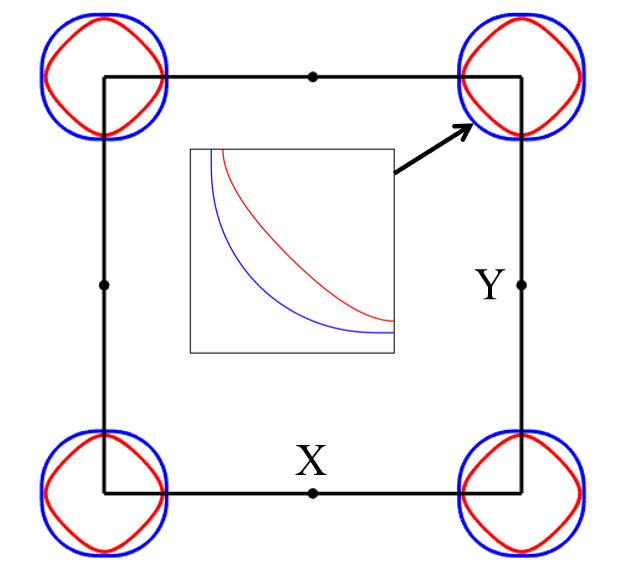}
	\caption{\label{fig_iron} (color online) The typical Fermi surfaces in the monolayer FeSe or LiFe(OH)FeSe in the presence of the spin-orbit coupling, with the details shown in the inset.   A sign reverse between the superconducting orders on the inner and outer Fermi surfaces, which  leads to the topological superconductivity as shown in Fig.\ref{fig2}(a),  are indicated by the red and blue colors in the figure.
	}
\end{figure}


\textit{Implication to the superconducting state in iron-based superconductors.}
In iron-based superconductors, in particular, in those iron chalcogenides, such as KFe$_2$Se$_2$\cite{zhang2011nodeless, PhysRevLett.106.187001}, the monolayer FeSe/STO\cite{liu2012electronic}, and LiFe(OH)FeSe\cite{zhao2016common}, where their Fermi surfaces are featured by two electronic pockets at the Brillioun zone corners as shown in Fig.\ref{fig_iron}(a),  there is a long-lasting debate about the pairing symmetry in their superconducting states\cite{Hirschfeld_2011, RevModPhys.87.855, doi:10.1146/annurev-conmatphys-031016-025242, RevModPhys.85.849}.  High-resolution angle-resolved photoemission spectroscopy and scanning tunneling spectroscopy (STM) measurements reveal a nodeless superconducting gap\cite{zhang2011nodeless, PhysRevLett.106.187001, liu2012electronic, zhao2016common, Wang_2012} in these materials. Although the results seem to be consistent with a conventional s-wave state\cite{PhysRevX.1.011009}, it has also been argued that  sign-changed s-wave state\cite{PhysRevB.84.024529, PhysRevB.85.054505, yin2014spin} between the inner and outer pockets can be  candidate as well. In the presence of strong spin-orbit coupling or inter-band pairing\cite{PhysRevB.84.024529, PhysRevX.3.031004, hu2013mechanism, PhysRevB.89.045144}, the latter  can be nodeless.  The experimental results  on this issue remain highly controversial with supporting evidence for both cases.  Indirect experimental evidence from inelastic neutron scattering\cite{RevModPhys.87.855} and STM quasiparticle interference\cite{du2018sign} exists for the sign changed s-wave while impurity scattering suggests the  conventional s-wave state\cite{fan2015plain,PhysRevX.1.011009}.

The Fermi surfaces in Fig.\ref{fig_iron}(a) are qualitatively the same as that in Fig.\ref{fig2}(a). Thus, if the iron chalcogenides are in the sign-changed s-wave states, we can expect  that they are second-order topological superconductors and must exhibit  two Dirac cones on the $(10)/(01)$ boundary and one Majorana Kramers' pair at the intersection between the $(11)$ and $(1 \overline{1} )$ edges.  To demonstrate this, we carry out our calculation in a five-orbital model  which describes the genuine band structures of iron-based superconductors\cite{wu2016nematic}. The result is given in the SM, which confirms that  iron chalcogenides, if they host the sign changed s-wave states between two electron pockets, are second-order topological superconductors.

This results offers smoking-gun evidence to reveal the pairing nature of iron-based superconductors. The monolayer FeSe and LiFe(OH)FeSe  are the two best systems for such a purpose as their superconducting gaps can be more than 10 meV. If they are in the sign-changed s-wave state,  the local probe, such as STM, can easily capture the boundary-selective Dirac cones  on the $(10)/(11)$ boundaries.  A tunnelling measurement at the intersection between the $(11)$ and $(1 \overline{1} )$ edge can measure the quantized zero-bias conductance peak of $4 e^2 / h$  resulted from  Majorana Kramers' pairs\cite{PhysRevLett.111.056402}.

\textit{Conclusions.}
In summary, we identify that the centrosymmetric superconductor, which respects nonsymmorphic symmetry group, can host topological superconductivity even though its superconducting order belongs to an s-wave state. The anomalous band degeneracies stemming from the nonsymmorphic symmetries are vital in realizing the topological superconductivity. The study is specified with the space group $P4/nmm$, with a Fermi surface criterion developed for the topological suerconductivity. Based on the analysis, we reveal that the iron-chalcogenide superconductors, if they are in the  sign-changed s-wave state,  are second-order topological superconductors,  which have two Dirac cones on the $(10)/(01)$ boundary and one Majorana Kramers' pair at the intersection between the $(11)$ and $(1 \overline{1} )$ edges. Our study  uncovers a new direction in the pursuit of the topological superconductors, and  establish a measurable quantity to  settle one long-lasting debates on the pairing nature of iron-based superconductors.

The authors are grateful to Xianxin Wu and Zhongyi Zhang for fruitful discussions. This work is supported by the Ministry of Science and Technology of China 973 program (Grant
No. 2017YFA0303100), National Science Foundation of China (Grant No. NSFC-11888101), and  the Strategic Priority Research Program of Chinese Academy of Sciences (XDB33000000).

%
%
%

\bibliographystyle{apsrev4-1}
\bibliography{reference}

\begin{thebibliography}{97}%
\makeatletter
\providecommand \@ifxundefined [1]{%
 \@ifx{#1\undefined}
}%
\providecommand \@ifnum [1]{%
 \ifnum #1\expandafter \@firstoftwo
 \else \expandafter \@secondoftwo
 \fi
}%
\providecommand \@ifx [1]{%
 \ifx #1\expandafter \@firstoftwo
 \else \expandafter \@secondoftwo
 \fi
}%
\providecommand \natexlab [1]{#1}%
\providecommand \enquote  [1]{``#1''}%
\providecommand \bibnamefont  [1]{#1}%
\providecommand \bibfnamefont [1]{#1}%
\providecommand \citenamefont [1]{#1}%
\providecommand \href@noop [0]{\@secondoftwo}%
\providecommand \href [0]{\begingroup \@sanitize@url \@href}%
\providecommand \@href[1]{\@@startlink{#1}\@@href}%
\providecommand \@@href[1]{\endgroup#1\@@endlink}%
\providecommand \@sanitize@url [0]{\catcode `\\12\catcode `\$12\catcode
  `\&12\catcode `\#12\catcode `\^12\catcode `\_12\catcode `\%12\relax}%
\providecommand \@@startlink[1]{}%
\providecommand \@@endlink[0]{}%
\providecommand \url  [0]{\begingroup\@sanitize@url \@url }%
\providecommand \@url [1]{\endgroup\@href {#1}{\urlprefix }}%
\providecommand \urlprefix  [0]{URL }%
\providecommand \Eprint [0]{\href }%
\providecommand \doibase [0]{http://dx.doi.org/}%
\providecommand \selectlanguage [0]{\@gobble}%
\providecommand \bibinfo  [0]{\@secondoftwo}%
\providecommand \bibfield  [0]{\@secondoftwo}%
\providecommand \translation [1]{[#1]}%
\providecommand \BibitemOpen [0]{}%
\providecommand \bibitemStop [0]{}%
\providecommand \bibitemNoStop [0]{.\EOS\space}%
\providecommand \EOS [0]{\spacefactor3000\relax}%
\providecommand \BibitemShut  [1]{\csname bibitem#1\endcsname}%
\let\auto@bib@innerbib\@empty
\bibitem [{\citenamefont {Nayak}\ \emph {et~al.}(2008)\citenamefont {Nayak},
  \citenamefont {Simon}, \citenamefont {Stern}, \citenamefont {Freedman},\ and\
  \citenamefont {Das~Sarma}}]{RevModPhys.80.1083}%
  \BibitemOpen
  \bibfield  {author} {\bibinfo {author} {\bibfnamefont {C.}~\bibnamefont
  {Nayak}}, \bibinfo {author} {\bibfnamefont {S.~H.}\ \bibnamefont {Simon}},
  \bibinfo {author} {\bibfnamefont {A.}~\bibnamefont {Stern}}, \bibinfo
  {author} {\bibfnamefont {M.}~\bibnamefont {Freedman}}, \ and\ \bibinfo
  {author} {\bibfnamefont {S.}~\bibnamefont {Das~Sarma}},\ }\href {\doibase
  10.1103/RevModPhys.80.1083} {\bibfield  {journal} {\bibinfo  {journal} {Rev.
  Mod. Phys.}\ }\textbf {\bibinfo {volume} {80}},\ \bibinfo {pages} {1083}
  (\bibinfo {year} {2008})}\BibitemShut {NoStop}%
\bibitem [{\citenamefont {Hasan}\ and\ \citenamefont
  {Kane}(2010)}]{RevModPhys.82.3045}%
  \BibitemOpen
  \bibfield  {author} {\bibinfo {author} {\bibfnamefont {M.~Z.}\ \bibnamefont
  {Hasan}}\ and\ \bibinfo {author} {\bibfnamefont {C.~L.}\ \bibnamefont
  {Kane}},\ }\href {\doibase 10.1103/RevModPhys.82.3045} {\bibfield  {journal}
  {\bibinfo  {journal} {Rev. Mod. Phys.}\ }\textbf {\bibinfo {volume} {82}},\
  \bibinfo {pages} {3045} (\bibinfo {year} {2010})}\BibitemShut {NoStop}%
\bibitem [{\citenamefont {Qi}\ and\ \citenamefont
  {Zhang}(2011)}]{RevModPhys.83.1057}%
  \BibitemOpen
  \bibfield  {author} {\bibinfo {author} {\bibfnamefont {X.-L.}\ \bibnamefont
  {Qi}}\ and\ \bibinfo {author} {\bibfnamefont {S.-C.}\ \bibnamefont {Zhang}},\
  }\href {\doibase 10.1103/RevModPhys.83.1057} {\bibfield  {journal} {\bibinfo
  {journal} {Rev. Mod. Phys.}\ }\textbf {\bibinfo {volume} {83}},\ \bibinfo
  {pages} {1057} (\bibinfo {year} {2011})}\BibitemShut {NoStop}%
\bibitem [{\citenamefont {Chiu}\ \emph {et~al.}(2016)\citenamefont {Chiu},
  \citenamefont {Teo}, \citenamefont {Schnyder},\ and\ \citenamefont
  {Ryu}}]{RevModPhys.88.035005}%
  \BibitemOpen
  \bibfield  {author} {\bibinfo {author} {\bibfnamefont {C.-K.}\ \bibnamefont
  {Chiu}}, \bibinfo {author} {\bibfnamefont {J.~C.~Y.}\ \bibnamefont {Teo}},
  \bibinfo {author} {\bibfnamefont {A.~P.}\ \bibnamefont {Schnyder}}, \ and\
  \bibinfo {author} {\bibfnamefont {S.}~\bibnamefont {Ryu}},\ }\href {\doibase
  10.1103/RevModPhys.88.035005} {\bibfield  {journal} {\bibinfo  {journal}
  {Rev. Mod. Phys.}\ }\textbf {\bibinfo {volume} {88}},\ \bibinfo {pages}
  {035005} (\bibinfo {year} {2016})}\BibitemShut {NoStop}%
\bibitem [{\citenamefont {Kitaev}(2001)}]{Kitaev_2001}%
  \BibitemOpen
  \bibfield  {author} {\bibinfo {author} {\bibfnamefont {A.~Y.}\ \bibnamefont
  {Kitaev}},\ }\href {\doibase 10.1070/1063-7869/44/10s/s29} {\bibfield
  {journal} {\bibinfo  {journal} {Physics-Uspekhi}\ }\textbf {\bibinfo {volume}
  {44}},\ \bibinfo {pages} {131} (\bibinfo {year} {2001})}\BibitemShut
  {NoStop}%
\bibitem [{\citenamefont {Alicea}(2012)}]{Alicea_2012}%
  \BibitemOpen
  \bibfield  {author} {\bibinfo {author} {\bibfnamefont {J.}~\bibnamefont
  {Alicea}},\ }\href {\doibase 10.1088/0034-4885/75/7/076501} {\bibfield
  {journal} {\bibinfo  {journal} {Reports on Progress in Physics}\ }\textbf
  {\bibinfo {volume} {75}},\ \bibinfo {pages} {076501} (\bibinfo {year}
  {2012})}\BibitemShut {NoStop}%
\bibitem [{\citenamefont {Hao}\ and\ \citenamefont
  {Hu}(2019)}]{hao2019topological}%
  \BibitemOpen
  \bibfield  {author} {\bibinfo {author} {\bibfnamefont {N.}~\bibnamefont
  {Hao}}\ and\ \bibinfo {author} {\bibfnamefont {J.}~\bibnamefont {Hu}},\
  }\href@noop {} {\bibfield  {journal} {\bibinfo  {journal} {National Science
  Review}\ }\textbf {\bibinfo {volume} {6}},\ \bibinfo {pages} {213} (\bibinfo
  {year} {2019})}\BibitemShut {NoStop}%
\bibitem [{\citenamefont {Wu}\ \emph {et~al.}(2020{\natexlab{a}})\citenamefont
  {Wu}, \citenamefont {Zhang}, \citenamefont {Xu}, \citenamefont {Hu},\ and\
  \citenamefont {Liu}}]{wu2020pursuit}%
  \BibitemOpen
  \bibfield  {author} {\bibinfo {author} {\bibfnamefont {X.}~\bibnamefont
  {Wu}}, \bibinfo {author} {\bibfnamefont {R.-X.}\ \bibnamefont {Zhang}},
  \bibinfo {author} {\bibfnamefont {G.}~\bibnamefont {Xu}}, \bibinfo {author}
  {\bibfnamefont {J.}~\bibnamefont {Hu}}, \ and\ \bibinfo {author}
  {\bibfnamefont {C.-X.}\ \bibnamefont {Liu}},\ }\href@noop {} {\bibfield
  {journal} {\bibinfo  {journal} {arXiv preprint arXiv:2005.03603}\ } (\bibinfo
  {year} {2020}{\natexlab{a}})}\BibitemShut {NoStop}%
\bibitem [{\citenamefont {Fu}\ and\ \citenamefont
  {Kane}(2008)}]{PhysRevLett.100.096407}%
  \BibitemOpen
  \bibfield  {author} {\bibinfo {author} {\bibfnamefont {L.}~\bibnamefont
  {Fu}}\ and\ \bibinfo {author} {\bibfnamefont {C.~L.}\ \bibnamefont {Kane}},\
  }\href {\doibase 10.1103/PhysRevLett.100.096407} {\bibfield  {journal}
  {\bibinfo  {journal} {Phys. Rev. Lett.}\ }\textbf {\bibinfo {volume} {100}},\
  \bibinfo {pages} {096407} (\bibinfo {year} {2008})}\BibitemShut {NoStop}%
\bibitem [{\citenamefont {Sau}\ \emph {et~al.}(2010)\citenamefont {Sau},
  \citenamefont {Lutchyn}, \citenamefont {Tewari},\ and\ \citenamefont
  {Das~Sarma}}]{PhysRevLett.104.040502}%
  \BibitemOpen
  \bibfield  {author} {\bibinfo {author} {\bibfnamefont {J.~D.}\ \bibnamefont
  {Sau}}, \bibinfo {author} {\bibfnamefont {R.~M.}\ \bibnamefont {Lutchyn}},
  \bibinfo {author} {\bibfnamefont {S.}~\bibnamefont {Tewari}}, \ and\ \bibinfo
  {author} {\bibfnamefont {S.}~\bibnamefont {Das~Sarma}},\ }\href {\doibase
  10.1103/PhysRevLett.104.040502} {\bibfield  {journal} {\bibinfo  {journal}
  {Phys. Rev. Lett.}\ }\textbf {\bibinfo {volume} {104}},\ \bibinfo {pages}
  {040502} (\bibinfo {year} {2010})}\BibitemShut {NoStop}%
\bibitem [{\citenamefont {Lutchyn}\ \emph {et~al.}(2010)\citenamefont
  {Lutchyn}, \citenamefont {Sau},\ and\ \citenamefont
  {Das~Sarma}}]{PhysRevLett.105.077001}%
  \BibitemOpen
  \bibfield  {author} {\bibinfo {author} {\bibfnamefont {R.~M.}\ \bibnamefont
  {Lutchyn}}, \bibinfo {author} {\bibfnamefont {J.~D.}\ \bibnamefont {Sau}}, \
  and\ \bibinfo {author} {\bibfnamefont {S.}~\bibnamefont {Das~Sarma}},\ }\href
  {\doibase 10.1103/PhysRevLett.105.077001} {\bibfield  {journal} {\bibinfo
  {journal} {Phys. Rev. Lett.}\ }\textbf {\bibinfo {volume} {105}},\ \bibinfo
  {pages} {077001} (\bibinfo {year} {2010})}\BibitemShut {NoStop}%
\bibitem [{\citenamefont {Oreg}\ \emph {et~al.}(2010)\citenamefont {Oreg},
  \citenamefont {Refael},\ and\ \citenamefont {von
  Oppen}}]{PhysRevLett.105.177002}%
  \BibitemOpen
  \bibfield  {author} {\bibinfo {author} {\bibfnamefont {Y.}~\bibnamefont
  {Oreg}}, \bibinfo {author} {\bibfnamefont {G.}~\bibnamefont {Refael}}, \ and\
  \bibinfo {author} {\bibfnamefont {F.}~\bibnamefont {von Oppen}},\ }\href
  {\doibase 10.1103/PhysRevLett.105.177002} {\bibfield  {journal} {\bibinfo
  {journal} {Phys. Rev. Lett.}\ }\textbf {\bibinfo {volume} {105}},\ \bibinfo
  {pages} {177002} (\bibinfo {year} {2010})}\BibitemShut {NoStop}%
\bibitem [{\citenamefont {Wimmer}\ \emph {et~al.}(2010)\citenamefont {Wimmer},
  \citenamefont {Akhmerov}, \citenamefont {Medvedyeva}, \citenamefont
  {Tworzyd\l{}o},\ and\ \citenamefont {Beenakker}}]{PhysRevLett.105.046803}%
  \BibitemOpen
  \bibfield  {author} {\bibinfo {author} {\bibfnamefont {M.}~\bibnamefont
  {Wimmer}}, \bibinfo {author} {\bibfnamefont {A.~R.}\ \bibnamefont
  {Akhmerov}}, \bibinfo {author} {\bibfnamefont {M.~V.}\ \bibnamefont
  {Medvedyeva}}, \bibinfo {author} {\bibfnamefont {J.}~\bibnamefont
  {Tworzyd\l{}o}}, \ and\ \bibinfo {author} {\bibfnamefont {C.~W.~J.}\
  \bibnamefont {Beenakker}},\ }\href {\doibase 10.1103/PhysRevLett.105.046803}
  {\bibfield  {journal} {\bibinfo  {journal} {Phys. Rev. Lett.}\ }\textbf
  {\bibinfo {volume} {105}},\ \bibinfo {pages} {046803} (\bibinfo {year}
  {2010})}\BibitemShut {NoStop}%
\bibitem [{\citenamefont {Teo}\ and\ \citenamefont
  {Kane}(2010)}]{PhysRevB.82.115120}%
  \BibitemOpen
  \bibfield  {author} {\bibinfo {author} {\bibfnamefont {J.~C.~Y.}\
  \bibnamefont {Teo}}\ and\ \bibinfo {author} {\bibfnamefont {C.~L.}\
  \bibnamefont {Kane}},\ }\href {\doibase 10.1103/PhysRevB.82.115120}
  {\bibfield  {journal} {\bibinfo  {journal} {Phys. Rev. B}\ }\textbf {\bibinfo
  {volume} {82}},\ \bibinfo {pages} {115120} (\bibinfo {year}
  {2010})}\BibitemShut {NoStop}%
\bibitem [{\citenamefont {Hosur}\ \emph {et~al.}(2011)\citenamefont {Hosur},
  \citenamefont {Ghaemi}, \citenamefont {Mong},\ and\ \citenamefont
  {Vishwanath}}]{PhysRevLett.107.097001}%
  \BibitemOpen
  \bibfield  {author} {\bibinfo {author} {\bibfnamefont {P.}~\bibnamefont
  {Hosur}}, \bibinfo {author} {\bibfnamefont {P.}~\bibnamefont {Ghaemi}},
  \bibinfo {author} {\bibfnamefont {R.~S.~K.}\ \bibnamefont {Mong}}, \ and\
  \bibinfo {author} {\bibfnamefont {A.}~\bibnamefont {Vishwanath}},\ }\href
  {\doibase 10.1103/PhysRevLett.107.097001} {\bibfield  {journal} {\bibinfo
  {journal} {Phys. Rev. Lett.}\ }\textbf {\bibinfo {volume} {107}},\ \bibinfo
  {pages} {097001} (\bibinfo {year} {2011})}\BibitemShut {NoStop}%
\bibitem [{\citenamefont {Teo}\ and\ \citenamefont
  {Hughes}(2013)}]{PhysRevLett.111.047006}%
  \BibitemOpen
  \bibfield  {author} {\bibinfo {author} {\bibfnamefont {J.~C.~Y.}\
  \bibnamefont {Teo}}\ and\ \bibinfo {author} {\bibfnamefont {T.~L.}\
  \bibnamefont {Hughes}},\ }\href {\doibase 10.1103/PhysRevLett.111.047006}
  {\bibfield  {journal} {\bibinfo  {journal} {Phys. Rev. Lett.}\ }\textbf
  {\bibinfo {volume} {111}},\ \bibinfo {pages} {047006} (\bibinfo {year}
  {2013})}\BibitemShut {NoStop}%
\bibitem [{\citenamefont {Zhang}\ \emph
  {et~al.}(2013{\natexlab{a}})\citenamefont {Zhang}, \citenamefont {Kane},\
  and\ \citenamefont {Mele}}]{PhysRevLett.111.056402}%
  \BibitemOpen
  \bibfield  {author} {\bibinfo {author} {\bibfnamefont {F.}~\bibnamefont
  {Zhang}}, \bibinfo {author} {\bibfnamefont {C.~L.}\ \bibnamefont {Kane}}, \
  and\ \bibinfo {author} {\bibfnamefont {E.~J.}\ \bibnamefont {Mele}},\ }\href
  {\doibase 10.1103/PhysRevLett.111.056402} {\bibfield  {journal} {\bibinfo
  {journal} {Phys. Rev. Lett.}\ }\textbf {\bibinfo {volume} {111}},\ \bibinfo
  {pages} {056402} (\bibinfo {year} {2013}{\natexlab{a}})}\BibitemShut
  {NoStop}%
\bibitem [{\citenamefont {Pientka}\ \emph {et~al.}(2013)\citenamefont
  {Pientka}, \citenamefont {Glazman},\ and\ \citenamefont {von
  Oppen}}]{PhysRevB.88.155420}%
  \BibitemOpen
  \bibfield  {author} {\bibinfo {author} {\bibfnamefont {F.}~\bibnamefont
  {Pientka}}, \bibinfo {author} {\bibfnamefont {L.~I.}\ \bibnamefont
  {Glazman}}, \ and\ \bibinfo {author} {\bibfnamefont {F.}~\bibnamefont {von
  Oppen}},\ }\href {\doibase 10.1103/PhysRevB.88.155420} {\bibfield  {journal}
  {\bibinfo  {journal} {Phys. Rev. B}\ }\textbf {\bibinfo {volume} {88}},\
  \bibinfo {pages} {155420} (\bibinfo {year} {2013})}\BibitemShut {NoStop}%
\bibitem [{\citenamefont {Sau}\ and\ \citenamefont
  {Brydon}(2015)}]{PhysRevLett.115.127003}%
  \BibitemOpen
  \bibfield  {author} {\bibinfo {author} {\bibfnamefont {J.~D.}\ \bibnamefont
  {Sau}}\ and\ \bibinfo {author} {\bibfnamefont {P.~M.~R.}\ \bibnamefont
  {Brydon}},\ }\href {\doibase 10.1103/PhysRevLett.115.127003} {\bibfield
  {journal} {\bibinfo  {journal} {Phys. Rev. Lett.}\ }\textbf {\bibinfo
  {volume} {115}},\ \bibinfo {pages} {127003} (\bibinfo {year}
  {2015})}\BibitemShut {NoStop}%
\bibitem [{\citenamefont {Li}\ \emph {et~al.}(2016)\citenamefont {Li},
  \citenamefont {Neupert}, \citenamefont {Bernevig},\ and\ \citenamefont
  {Yazdani}}]{li2016manipulating}%
  \BibitemOpen
  \bibfield  {author} {\bibinfo {author} {\bibfnamefont {J.}~\bibnamefont
  {Li}}, \bibinfo {author} {\bibfnamefont {T.}~\bibnamefont {Neupert}},
  \bibinfo {author} {\bibfnamefont {B.~A.}\ \bibnamefont {Bernevig}}, \ and\
  \bibinfo {author} {\bibfnamefont {A.}~\bibnamefont {Yazdani}},\ }\href@noop
  {} {\bibfield  {journal} {\bibinfo  {journal} {Nature communications}\
  }\textbf {\bibinfo {volume} {7}},\ \bibinfo {pages} {1} (\bibinfo {year}
  {2016})}\BibitemShut {NoStop}%
\bibitem [{\citenamefont {Xu}\ \emph {et~al.}(2016)\citenamefont {Xu},
  \citenamefont {Lian}, \citenamefont {Tang}, \citenamefont {Qi},\ and\
  \citenamefont {Zhang}}]{PhysRevLett.117.047001}%
  \BibitemOpen
  \bibfield  {author} {\bibinfo {author} {\bibfnamefont {G.}~\bibnamefont
  {Xu}}, \bibinfo {author} {\bibfnamefont {B.}~\bibnamefont {Lian}}, \bibinfo
  {author} {\bibfnamefont {P.}~\bibnamefont {Tang}}, \bibinfo {author}
  {\bibfnamefont {X.-L.}\ \bibnamefont {Qi}}, \ and\ \bibinfo {author}
  {\bibfnamefont {S.-C.}\ \bibnamefont {Zhang}},\ }\href {\doibase
  10.1103/PhysRevLett.117.047001} {\bibfield  {journal} {\bibinfo  {journal}
  {Phys. Rev. Lett.}\ }\textbf {\bibinfo {volume} {117}},\ \bibinfo {pages}
  {047001} (\bibinfo {year} {2016})}\BibitemShut {NoStop}%
\bibitem [{\citenamefont {Yang}\ \emph {et~al.}(2016)\citenamefont {Yang},
  \citenamefont {Stano}, \citenamefont {Klinovaja},\ and\ \citenamefont
  {Loss}}]{yang2016majorana}%
  \BibitemOpen
  \bibfield  {author} {\bibinfo {author} {\bibfnamefont {G.}~\bibnamefont
  {Yang}}, \bibinfo {author} {\bibfnamefont {P.}~\bibnamefont {Stano}},
  \bibinfo {author} {\bibfnamefont {J.}~\bibnamefont {Klinovaja}}, \ and\
  \bibinfo {author} {\bibfnamefont {D.}~\bibnamefont {Loss}},\ }\href@noop {}
  {\bibfield  {journal} {\bibinfo  {journal} {Physical Review B}\ }\textbf
  {\bibinfo {volume} {93}},\ \bibinfo {pages} {224505} (\bibinfo {year}
  {2016})}\BibitemShut {NoStop}%
\bibitem [{\citenamefont {Jiang}\ \emph {et~al.}(2019)\citenamefont {Jiang},
  \citenamefont {Dai},\ and\ \citenamefont {Wang}}]{PhysRevX.9.011033}%
  \BibitemOpen
  \bibfield  {author} {\bibinfo {author} {\bibfnamefont {K.}~\bibnamefont
  {Jiang}}, \bibinfo {author} {\bibfnamefont {X.}~\bibnamefont {Dai}}, \ and\
  \bibinfo {author} {\bibfnamefont {Z.}~\bibnamefont {Wang}},\ }\href {\doibase
  10.1103/PhysRevX.9.011033} {\bibfield  {journal} {\bibinfo  {journal} {Phys.
  Rev. X}\ }\textbf {\bibinfo {volume} {9}},\ \bibinfo {pages} {011033}
  (\bibinfo {year} {2019})}\BibitemShut {NoStop}%
\bibitem [{\citenamefont {Sasaki}\ \emph {et~al.}(2011)\citenamefont {Sasaki},
  \citenamefont {Kriener}, \citenamefont {Segawa}, \citenamefont {Yada},
  \citenamefont {Tanaka}, \citenamefont {Sato},\ and\ \citenamefont
  {Ando}}]{TSC_Ando}%
  \BibitemOpen
  \bibfield  {author} {\bibinfo {author} {\bibfnamefont {S.}~\bibnamefont
  {Sasaki}}, \bibinfo {author} {\bibfnamefont {M.}~\bibnamefont {Kriener}},
  \bibinfo {author} {\bibfnamefont {K.}~\bibnamefont {Segawa}}, \bibinfo
  {author} {\bibfnamefont {K.}~\bibnamefont {Yada}}, \bibinfo {author}
  {\bibfnamefont {Y.}~\bibnamefont {Tanaka}}, \bibinfo {author} {\bibfnamefont
  {M.}~\bibnamefont {Sato}}, \ and\ \bibinfo {author} {\bibfnamefont
  {Y.}~\bibnamefont {Ando}},\ }\href {\doibase 10.1103/PhysRevLett.107.217001}
  {\bibfield  {journal} {\bibinfo  {journal} {Phys. Rev. Lett.}\ }\textbf
  {\bibinfo {volume} {107}},\ \bibinfo {pages} {217001} (\bibinfo {year}
  {2011})}\BibitemShut {NoStop}%
\bibitem [{\citenamefont {Das}\ \emph {et~al.}(2012)\citenamefont {Das},
  \citenamefont {Ronen}, \citenamefont {Most}, \citenamefont {Oreg},
  \citenamefont {Heiblum},\ and\ \citenamefont {Shtrikman}}]{das2012zero}%
  \BibitemOpen
  \bibfield  {author} {\bibinfo {author} {\bibfnamefont {A.}~\bibnamefont
  {Das}}, \bibinfo {author} {\bibfnamefont {Y.}~\bibnamefont {Ronen}}, \bibinfo
  {author} {\bibfnamefont {Y.}~\bibnamefont {Most}}, \bibinfo {author}
  {\bibfnamefont {Y.}~\bibnamefont {Oreg}}, \bibinfo {author} {\bibfnamefont
  {M.}~\bibnamefont {Heiblum}}, \ and\ \bibinfo {author} {\bibfnamefont
  {H.}~\bibnamefont {Shtrikman}},\ }\href@noop {} {\bibfield  {journal}
  {\bibinfo  {journal} {Nature Physics}\ }\textbf {\bibinfo {volume} {8}},\
  \bibinfo {pages} {887} (\bibinfo {year} {2012})}\BibitemShut {NoStop}%
\bibitem [{\citenamefont {Nadj-Perge}\ \emph {et~al.}(2014)\citenamefont
  {Nadj-Perge}, \citenamefont {Drozdov}, \citenamefont {Li}, \citenamefont
  {Chen}, \citenamefont {Jeon}, \citenamefont {Seo}, \citenamefont {MacDonald},
  \citenamefont {Bernevig},\ and\ \citenamefont
  {Yazdani}}]{nadj2014observation}%
  \BibitemOpen
  \bibfield  {author} {\bibinfo {author} {\bibfnamefont {S.}~\bibnamefont
  {Nadj-Perge}}, \bibinfo {author} {\bibfnamefont {I.~K.}\ \bibnamefont
  {Drozdov}}, \bibinfo {author} {\bibfnamefont {J.}~\bibnamefont {Li}},
  \bibinfo {author} {\bibfnamefont {H.}~\bibnamefont {Chen}}, \bibinfo {author}
  {\bibfnamefont {S.}~\bibnamefont {Jeon}}, \bibinfo {author} {\bibfnamefont
  {J.}~\bibnamefont {Seo}}, \bibinfo {author} {\bibfnamefont {A.~H.}\
  \bibnamefont {MacDonald}}, \bibinfo {author} {\bibfnamefont {B.~A.}\
  \bibnamefont {Bernevig}}, \ and\ \bibinfo {author} {\bibfnamefont
  {A.}~\bibnamefont {Yazdani}},\ }\href {\doibase 10.1126/science.1259327}
  {\bibfield  {journal} {\bibinfo  {journal} {Science}\ }\textbf {\bibinfo
  {volume} {346}} (\bibinfo {year} {2014}),\
  10.1126/science.1259327}\BibitemShut {NoStop}%
\bibitem [{\citenamefont {Xu}\ \emph {et~al.}(2015)\citenamefont {Xu},
  \citenamefont {Wang}, \citenamefont {Liu}, \citenamefont {Ge}, \citenamefont
  {Yang}, \citenamefont {Liu}, \citenamefont {Xu}, \citenamefont {Guan},
  \citenamefont {Gao}, \citenamefont {Qian}, \citenamefont {Liu}, \citenamefont
  {Wang}, \citenamefont {Zhang}, \citenamefont {Xue},\ and\ \citenamefont
  {Jia}}]{MZM_Jia}%
  \BibitemOpen
  \bibfield  {author} {\bibinfo {author} {\bibfnamefont {J.-P.}\ \bibnamefont
  {Xu}}, \bibinfo {author} {\bibfnamefont {M.-X.}\ \bibnamefont {Wang}},
  \bibinfo {author} {\bibfnamefont {Z.~L.}\ \bibnamefont {Liu}}, \bibinfo
  {author} {\bibfnamefont {J.-F.}\ \bibnamefont {Ge}}, \bibinfo {author}
  {\bibfnamefont {X.}~\bibnamefont {Yang}}, \bibinfo {author} {\bibfnamefont
  {C.}~\bibnamefont {Liu}}, \bibinfo {author} {\bibfnamefont {Z.~A.}\
  \bibnamefont {Xu}}, \bibinfo {author} {\bibfnamefont {D.}~\bibnamefont
  {Guan}}, \bibinfo {author} {\bibfnamefont {C.~L.}\ \bibnamefont {Gao}},
  \bibinfo {author} {\bibfnamefont {D.}~\bibnamefont {Qian}}, \bibinfo {author}
  {\bibfnamefont {Y.}~\bibnamefont {Liu}}, \bibinfo {author} {\bibfnamefont
  {Q.-H.}\ \bibnamefont {Wang}}, \bibinfo {author} {\bibfnamefont {F.-C.}\
  \bibnamefont {Zhang}}, \bibinfo {author} {\bibfnamefont {Q.-K.}\ \bibnamefont
  {Xue}}, \ and\ \bibinfo {author} {\bibfnamefont {J.-F.}\ \bibnamefont
  {Jia}},\ }\href {\doibase 10.1103/PhysRevLett.114.017001} {\bibfield
  {journal} {\bibinfo  {journal} {Phys. Rev. Lett.}\ }\textbf {\bibinfo
  {volume} {114}},\ \bibinfo {pages} {017001} (\bibinfo {year}
  {2015})}\BibitemShut {NoStop}%
\bibitem [{\citenamefont {Wang}\ \emph
  {et~al.}(2018{\natexlab{a}})\citenamefont {Wang}, \citenamefont {Kong},
  \citenamefont {Fan}, \citenamefont {Chen}, \citenamefont {Zhu}, \citenamefont
  {Liu}, \citenamefont {Cao}, \citenamefont {Sun}, \citenamefont {Du},
  \citenamefont {Schneeloch} \emph {et~al.}}]{wang2018evidence}%
  \BibitemOpen
  \bibfield  {author} {\bibinfo {author} {\bibfnamefont {D.}~\bibnamefont
  {Wang}}, \bibinfo {author} {\bibfnamefont {L.}~\bibnamefont {Kong}}, \bibinfo
  {author} {\bibfnamefont {P.}~\bibnamefont {Fan}}, \bibinfo {author}
  {\bibfnamefont {H.}~\bibnamefont {Chen}}, \bibinfo {author} {\bibfnamefont
  {S.}~\bibnamefont {Zhu}}, \bibinfo {author} {\bibfnamefont {W.}~\bibnamefont
  {Liu}}, \bibinfo {author} {\bibfnamefont {L.}~\bibnamefont {Cao}}, \bibinfo
  {author} {\bibfnamefont {Y.}~\bibnamefont {Sun}}, \bibinfo {author}
  {\bibfnamefont {S.}~\bibnamefont {Du}}, \bibinfo {author} {\bibfnamefont
  {J.}~\bibnamefont {Schneeloch}},  \emph {et~al.},\ }\href {\doibase
  10.1126/science.aao1797} {\bibfield  {journal} {\bibinfo  {journal}
  {Science}\ }\textbf {\bibinfo {volume} {362}} (\bibinfo {year}
  {2018}{\natexlab{a}}),\ 10.1126/science.aao1797}\BibitemShut {NoStop}%
\bibitem [{\citenamefont {Liu}\ \emph {et~al.}(2018)\citenamefont {Liu},
  \citenamefont {Chen}, \citenamefont {Zhang}, \citenamefont {Peng},
  \citenamefont {Yan}, \citenamefont {Wen}, \citenamefont {Lou}, \citenamefont
  {Huang}, \citenamefont {Tian}, \citenamefont {Dong}, \citenamefont {Wang},
  \citenamefont {Bao}, \citenamefont {Wang}, \citenamefont {Yin}, \citenamefont
  {Zhao},\ and\ \citenamefont {Feng}}]{vortex_Feng}%
  \BibitemOpen
  \bibfield  {author} {\bibinfo {author} {\bibfnamefont {Q.}~\bibnamefont
  {Liu}}, \bibinfo {author} {\bibfnamefont {C.}~\bibnamefont {Chen}}, \bibinfo
  {author} {\bibfnamefont {T.}~\bibnamefont {Zhang}}, \bibinfo {author}
  {\bibfnamefont {R.}~\bibnamefont {Peng}}, \bibinfo {author} {\bibfnamefont
  {Y.-J.}\ \bibnamefont {Yan}}, \bibinfo {author} {\bibfnamefont {C.-H.-P.}\
  \bibnamefont {Wen}}, \bibinfo {author} {\bibfnamefont {X.}~\bibnamefont
  {Lou}}, \bibinfo {author} {\bibfnamefont {Y.-L.}\ \bibnamefont {Huang}},
  \bibinfo {author} {\bibfnamefont {J.-P.}\ \bibnamefont {Tian}}, \bibinfo
  {author} {\bibfnamefont {X.-L.}\ \bibnamefont {Dong}}, \bibinfo {author}
  {\bibfnamefont {G.-W.}\ \bibnamefont {Wang}}, \bibinfo {author}
  {\bibfnamefont {W.-C.}\ \bibnamefont {Bao}}, \bibinfo {author} {\bibfnamefont
  {Q.-H.}\ \bibnamefont {Wang}}, \bibinfo {author} {\bibfnamefont {Z.-P.}\
  \bibnamefont {Yin}}, \bibinfo {author} {\bibfnamefont {Z.-X.}\ \bibnamefont
  {Zhao}}, \ and\ \bibinfo {author} {\bibfnamefont {D.-L.}\ \bibnamefont
  {Feng}},\ }\href {\doibase 10.1103/PhysRevX.8.041056} {\bibfield  {journal}
  {\bibinfo  {journal} {Phys. Rev. X}\ }\textbf {\bibinfo {volume} {8}},\
  \bibinfo {pages} {041056} (\bibinfo {year} {2018})}\BibitemShut {NoStop}%
\bibitem [{\citenamefont {Kong}\ \emph {et~al.}(2019)\citenamefont {Kong},
  \citenamefont {Zhu}, \citenamefont {Papaj}, \citenamefont {Chen},
  \citenamefont {Cao}, \citenamefont {Isobe}, \citenamefont {Xing},
  \citenamefont {Liu}, \citenamefont {Wang}, \citenamefont {Fan}, \citenamefont
  {Sun}, \citenamefont {Du}, \citenamefont {Schneeloch}, \citenamefont {Zhong},
  \citenamefont {Gu}, \citenamefont {Fu}, \citenamefont {Gao},\ and\
  \citenamefont {Ding}}]{vortex_Kong}%
  \BibitemOpen
  \bibfield  {author} {\bibinfo {author} {\bibfnamefont {L.}~\bibnamefont
  {Kong}}, \bibinfo {author} {\bibfnamefont {S.}~\bibnamefont {Zhu}}, \bibinfo
  {author} {\bibfnamefont {M.}~\bibnamefont {Papaj}}, \bibinfo {author}
  {\bibfnamefont {H.}~\bibnamefont {Chen}}, \bibinfo {author} {\bibfnamefont
  {L.}~\bibnamefont {Cao}}, \bibinfo {author} {\bibfnamefont {H.}~\bibnamefont
  {Isobe}}, \bibinfo {author} {\bibfnamefont {Y.}~\bibnamefont {Xing}},
  \bibinfo {author} {\bibfnamefont {W.}~\bibnamefont {Liu}}, \bibinfo {author}
  {\bibfnamefont {D.}~\bibnamefont {Wang}}, \bibinfo {author} {\bibfnamefont
  {P.}~\bibnamefont {Fan}}, \bibinfo {author} {\bibfnamefont {Y.}~\bibnamefont
  {Sun}}, \bibinfo {author} {\bibfnamefont {S.}~\bibnamefont {Du}}, \bibinfo
  {author} {\bibfnamefont {J.}~\bibnamefont {Schneeloch}}, \bibinfo {author}
  {\bibfnamefont {R.}~\bibnamefont {Zhong}}, \bibinfo {author} {\bibfnamefont
  {G.}~\bibnamefont {Gu}}, \bibinfo {author} {\bibfnamefont {L.}~\bibnamefont
  {Fu}}, \bibinfo {author} {\bibfnamefont {H.-J.}\ \bibnamefont {Gao}}, \ and\
  \bibinfo {author} {\bibfnamefont {H.}~\bibnamefont {Ding}},\ }\href {\doibase
  10.1038/s41567-019-0630-5} {\bibfield  {journal} {\bibinfo  {journal} {Nature
  Physics}\ }\textbf {\bibinfo {volume} {15}},\ \bibinfo {pages} {1181}
  (\bibinfo {year} {2019})}\BibitemShut {NoStop}%
\bibitem [{\citenamefont {Machida}\ \emph {et~al.}(2019)\citenamefont
  {Machida}, \citenamefont {Sun}, \citenamefont {Pyon}, \citenamefont {Takeda},
  \citenamefont {Kohsaka}, \citenamefont {Hanaguri}, \citenamefont {Sasagawa},\
  and\ \citenamefont {Tamegai}}]{machida2019zero}%
  \BibitemOpen
  \bibfield  {author} {\bibinfo {author} {\bibfnamefont {T.}~\bibnamefont
  {Machida}}, \bibinfo {author} {\bibfnamefont {Y.}~\bibnamefont {Sun}},
  \bibinfo {author} {\bibfnamefont {S.}~\bibnamefont {Pyon}}, \bibinfo {author}
  {\bibfnamefont {S.}~\bibnamefont {Takeda}}, \bibinfo {author} {\bibfnamefont
  {Y.}~\bibnamefont {Kohsaka}}, \bibinfo {author} {\bibfnamefont
  {T.}~\bibnamefont {Hanaguri}}, \bibinfo {author} {\bibfnamefont
  {T.}~\bibnamefont {Sasagawa}}, \ and\ \bibinfo {author} {\bibfnamefont
  {T.}~\bibnamefont {Tamegai}},\ }\href {\doibase 10.1038/s41563-019-0397-1}
  {\bibfield  {journal} {\bibinfo  {journal} {Nature materials}\ }\textbf
  {\bibinfo {volume} {18}},\ \bibinfo {pages} {811} (\bibinfo {year}
  {2019})}\BibitemShut {NoStop}%
\bibitem [{\citenamefont {Chen}\ \emph {et~al.}(2019)\citenamefont {Chen},
  \citenamefont {Chen}, \citenamefont {Duan}, \citenamefont {Zhu},
  \citenamefont {Yang},\ and\ \citenamefont {Wen}}]{chen2019observation}%
  \BibitemOpen
  \bibfield  {author} {\bibinfo {author} {\bibfnamefont {X.}~\bibnamefont
  {Chen}}, \bibinfo {author} {\bibfnamefont {M.}~\bibnamefont {Chen}}, \bibinfo
  {author} {\bibfnamefont {W.}~\bibnamefont {Duan}}, \bibinfo {author}
  {\bibfnamefont {X.}~\bibnamefont {Zhu}}, \bibinfo {author} {\bibfnamefont
  {H.}~\bibnamefont {Yang}}, \ and\ \bibinfo {author} {\bibfnamefont {H.-H.}\
  \bibnamefont {Wen}},\ }\href@noop {} {\bibfield  {journal} {\bibinfo
  {journal} {arXiv preprint arXiv:1909.01686}\ } (\bibinfo {year}
  {2019})}\BibitemShut {NoStop}%
\bibitem [{\citenamefont {Kong}\ \emph {et~al.}(2020)\citenamefont {Kong},
  \citenamefont {Cao}, \citenamefont {Zhu}, \citenamefont {Papaj},
  \citenamefont {Dai}, \citenamefont {Li}, \citenamefont {Fan}, \citenamefont
  {Liu}, \citenamefont {Yang}, \citenamefont {Wang} \emph
  {et~al.}}]{kong2020tunable}%
  \BibitemOpen
  \bibfield  {author} {\bibinfo {author} {\bibfnamefont {L.}~\bibnamefont
  {Kong}}, \bibinfo {author} {\bibfnamefont {L.}~\bibnamefont {Cao}}, \bibinfo
  {author} {\bibfnamefont {S.}~\bibnamefont {Zhu}}, \bibinfo {author}
  {\bibfnamefont {M.}~\bibnamefont {Papaj}}, \bibinfo {author} {\bibfnamefont
  {G.}~\bibnamefont {Dai}}, \bibinfo {author} {\bibfnamefont {G.}~\bibnamefont
  {Li}}, \bibinfo {author} {\bibfnamefont {P.}~\bibnamefont {Fan}}, \bibinfo
  {author} {\bibfnamefont {W.}~\bibnamefont {Liu}}, \bibinfo {author}
  {\bibfnamefont {F.}~\bibnamefont {Yang}}, \bibinfo {author} {\bibfnamefont
  {X.}~\bibnamefont {Wang}},  \emph {et~al.},\ }\href@noop {} {\bibfield
  {journal} {\bibinfo  {journal} {arXiv preprint arXiv:2010.04735}\ } (\bibinfo
  {year} {2020})}\BibitemShut {NoStop}%
\bibitem [{\citenamefont {Wang}\ \emph {et~al.}(2020)\citenamefont {Wang},
  \citenamefont {Rodriguez}, \citenamefont {Jiao}, \citenamefont {Howard},
  \citenamefont {Graham}, \citenamefont {Gu}, \citenamefont {Hughes},
  \citenamefont {Morr},\ and\ \citenamefont {Madhavan}}]{wang2020evidence}%
  \BibitemOpen
  \bibfield  {author} {\bibinfo {author} {\bibfnamefont {Z.}~\bibnamefont
  {Wang}}, \bibinfo {author} {\bibfnamefont {J.~O.}\ \bibnamefont {Rodriguez}},
  \bibinfo {author} {\bibfnamefont {L.}~\bibnamefont {Jiao}}, \bibinfo {author}
  {\bibfnamefont {S.}~\bibnamefont {Howard}}, \bibinfo {author} {\bibfnamefont
  {M.}~\bibnamefont {Graham}}, \bibinfo {author} {\bibfnamefont
  {G.}~\bibnamefont {Gu}}, \bibinfo {author} {\bibfnamefont {T.~L.}\
  \bibnamefont {Hughes}}, \bibinfo {author} {\bibfnamefont {D.~K.}\
  \bibnamefont {Morr}}, \ and\ \bibinfo {author} {\bibfnamefont
  {V.}~\bibnamefont {Madhavan}},\ }\href {\doibase 10.1126/science.aaw8419}
  {\bibfield  {journal} {\bibinfo  {journal} {Science}\ }\textbf {\bibinfo
  {volume} {367}} (\bibinfo {year} {2020}),\
  10.1126/science.aaw8419}\BibitemShut {NoStop}%
\bibitem [{\citenamefont {Vaitiekenas}\ \emph {et~al.}(2020)\citenamefont
  {Vaitiekenas}, \citenamefont {Winkler}, \citenamefont {van Heck},
  \citenamefont {Karzig}, \citenamefont {Deng}, \citenamefont {Flensberg},
  \citenamefont {Glazman}, \citenamefont {Nayak}, \citenamefont {Krogstrup},
  \citenamefont {Lutchyn},\ and\ \citenamefont {Marcus}}]{MZM_flux}%
  \BibitemOpen
  \bibfield  {author} {\bibinfo {author} {\bibfnamefont {S.}~\bibnamefont
  {Vaitiekenas}}, \bibinfo {author} {\bibfnamefont {G.~W.}\ \bibnamefont
  {Winkler}}, \bibinfo {author} {\bibfnamefont {B.}~\bibnamefont {van Heck}},
  \bibinfo {author} {\bibfnamefont {T.}~\bibnamefont {Karzig}}, \bibinfo
  {author} {\bibfnamefont {M.~T.}\ \bibnamefont {Deng}}, \bibinfo {author}
  {\bibfnamefont {K.}~\bibnamefont {Flensberg}}, \bibinfo {author}
  {\bibfnamefont {L.~I.}\ \bibnamefont {Glazman}}, \bibinfo {author}
  {\bibfnamefont {C.}~\bibnamefont {Nayak}}, \bibinfo {author} {\bibfnamefont
  {P.}~\bibnamefont {Krogstrup}}, \bibinfo {author} {\bibfnamefont {R.~M.}\
  \bibnamefont {Lutchyn}}, \ and\ \bibinfo {author} {\bibfnamefont {C.~M.}\
  \bibnamefont {Marcus}},\ }\href {\doibase 10.1126/science.aav3392} {\bibfield
   {journal} {\bibinfo  {journal} {Science}\ }\textbf {\bibinfo {volume} {367}}
  (\bibinfo {year} {2020}),\ 10.1126/science.aav3392}\BibitemShut {NoStop}%
\bibitem [{\citenamefont {Ando}\ and\ \citenamefont
  {Fu}(2015)}]{doi:10.1146/annurev-conmatphys-031214-014501}%
  \BibitemOpen
  \bibfield  {author} {\bibinfo {author} {\bibfnamefont {Y.}~\bibnamefont
  {Ando}}\ and\ \bibinfo {author} {\bibfnamefont {L.}~\bibnamefont {Fu}},\
  }\href {\doibase 10.1146/annurev-conmatphys-031214-014501} {\bibfield
  {journal} {\bibinfo  {journal} {Annual Review of Condensed Matter Physics}\
  }\textbf {\bibinfo {volume} {6}},\ \bibinfo {pages} {361} (\bibinfo {year}
  {2015})},\ \Eprint
  {http://arxiv.org/abs/https://doi.org/10.1146/annurev-conmatphys-031214-014501}
  {https://doi.org/10.1146/annurev-conmatphys-031214-014501} \BibitemShut
  {NoStop}%
\bibitem [{\citenamefont {Fu}(2011)}]{PhysRevLett.106.106802}%
  \BibitemOpen
  \bibfield  {author} {\bibinfo {author} {\bibfnamefont {L.}~\bibnamefont
  {Fu}},\ }\href {\doibase 10.1103/PhysRevLett.106.106802} {\bibfield
  {journal} {\bibinfo  {journal} {Phys. Rev. Lett.}\ }\textbf {\bibinfo
  {volume} {106}},\ \bibinfo {pages} {106802} (\bibinfo {year}
  {2011})}\BibitemShut {NoStop}%
\bibitem [{\citenamefont {Hsieh}\ \emph {et~al.}(2012)\citenamefont {Hsieh},
  \citenamefont {Lin}, \citenamefont {Liu}, \citenamefont {Duan}, \citenamefont
  {Bansil},\ and\ \citenamefont {Fu}}]{hsieh2012topological}%
  \BibitemOpen
  \bibfield  {author} {\bibinfo {author} {\bibfnamefont {T.~H.}\ \bibnamefont
  {Hsieh}}, \bibinfo {author} {\bibfnamefont {H.}~\bibnamefont {Lin}}, \bibinfo
  {author} {\bibfnamefont {J.}~\bibnamefont {Liu}}, \bibinfo {author}
  {\bibfnamefont {W.}~\bibnamefont {Duan}}, \bibinfo {author} {\bibfnamefont
  {A.}~\bibnamefont {Bansil}}, \ and\ \bibinfo {author} {\bibfnamefont
  {L.}~\bibnamefont {Fu}},\ }\href@noop {} {\bibfield  {journal} {\bibinfo
  {journal} {Nature communications}\ }\textbf {\bibinfo {volume} {3}},\
  \bibinfo {pages} {1} (\bibinfo {year} {2012})}\BibitemShut {NoStop}%
\bibitem [{\citenamefont {Tanaka}\ \emph {et~al.}(2012)\citenamefont {Tanaka},
  \citenamefont {Ren}, \citenamefont {Sato}, \citenamefont {Nakayama},
  \citenamefont {Souma}, \citenamefont {Takahashi}, \citenamefont {Segawa},\
  and\ \citenamefont {Ando}}]{tanaka2012experimental}%
  \BibitemOpen
  \bibfield  {author} {\bibinfo {author} {\bibfnamefont {Y.}~\bibnamefont
  {Tanaka}}, \bibinfo {author} {\bibfnamefont {Z.}~\bibnamefont {Ren}},
  \bibinfo {author} {\bibfnamefont {T.}~\bibnamefont {Sato}}, \bibinfo {author}
  {\bibfnamefont {K.}~\bibnamefont {Nakayama}}, \bibinfo {author}
  {\bibfnamefont {S.}~\bibnamefont {Souma}}, \bibinfo {author} {\bibfnamefont
  {T.}~\bibnamefont {Takahashi}}, \bibinfo {author} {\bibfnamefont
  {K.}~\bibnamefont {Segawa}}, \ and\ \bibinfo {author} {\bibfnamefont
  {Y.}~\bibnamefont {Ando}},\ }\href@noop {} {\bibfield  {journal} {\bibinfo
  {journal} {Nature Physics}\ }\textbf {\bibinfo {volume} {8}},\ \bibinfo
  {pages} {800} (\bibinfo {year} {2012})}\BibitemShut {NoStop}%
\bibitem [{\citenamefont {Bradlyn}\ \emph {et~al.}(2017)\citenamefont
  {Bradlyn}, \citenamefont {Elcoro}, \citenamefont {Cano}, \citenamefont
  {Vergniory}, \citenamefont {Wang}, \citenamefont {Felser}, \citenamefont
  {Aroyo},\ and\ \citenamefont {Bernevig}}]{bradlyn2017topological}%
  \BibitemOpen
  \bibfield  {author} {\bibinfo {author} {\bibfnamefont {B.}~\bibnamefont
  {Bradlyn}}, \bibinfo {author} {\bibfnamefont {L.}~\bibnamefont {Elcoro}},
  \bibinfo {author} {\bibfnamefont {J.}~\bibnamefont {Cano}}, \bibinfo {author}
  {\bibfnamefont {M.}~\bibnamefont {Vergniory}}, \bibinfo {author}
  {\bibfnamefont {Z.}~\bibnamefont {Wang}}, \bibinfo {author} {\bibfnamefont
  {C.}~\bibnamefont {Felser}}, \bibinfo {author} {\bibfnamefont
  {M.}~\bibnamefont {Aroyo}}, \ and\ \bibinfo {author} {\bibfnamefont {B.~A.}\
  \bibnamefont {Bernevig}},\ }\href@noop {} {\bibfield  {journal} {\bibinfo
  {journal} {Nature}\ }\textbf {\bibinfo {volume} {547}},\ \bibinfo {pages}
  {298} (\bibinfo {year} {2017})}\BibitemShut {NoStop}%
\bibitem [{\citenamefont {Po}\ \emph {et~al.}(2017)\citenamefont {Po},
  \citenamefont {Vishwanath},\ and\ \citenamefont {Watanabe}}]{po2017symmetry}%
  \BibitemOpen
  \bibfield  {author} {\bibinfo {author} {\bibfnamefont {H.~C.}\ \bibnamefont
  {Po}}, \bibinfo {author} {\bibfnamefont {A.}~\bibnamefont {Vishwanath}}, \
  and\ \bibinfo {author} {\bibfnamefont {H.}~\bibnamefont {Watanabe}},\
  }\href@noop {} {\bibfield  {journal} {\bibinfo  {journal} {Nature
  communications}\ }\textbf {\bibinfo {volume} {8}},\ \bibinfo {pages} {1}
  (\bibinfo {year} {2017})}\BibitemShut {NoStop}%
\bibitem [{\citenamefont {Kruthoff}\ \emph {et~al.}(2017)\citenamefont
  {Kruthoff}, \citenamefont {de~Boer}, \citenamefont {van Wezel}, \citenamefont
  {Kane},\ and\ \citenamefont {Slager}}]{PhysRevX.7.041069}%
  \BibitemOpen
  \bibfield  {author} {\bibinfo {author} {\bibfnamefont {J.}~\bibnamefont
  {Kruthoff}}, \bibinfo {author} {\bibfnamefont {J.}~\bibnamefont {de~Boer}},
  \bibinfo {author} {\bibfnamefont {J.}~\bibnamefont {van Wezel}}, \bibinfo
  {author} {\bibfnamefont {C.~L.}\ \bibnamefont {Kane}}, \ and\ \bibinfo
  {author} {\bibfnamefont {R.-J.}\ \bibnamefont {Slager}},\ }\href {\doibase
  10.1103/PhysRevX.7.041069} {\bibfield  {journal} {\bibinfo  {journal} {Phys.
  Rev. X}\ }\textbf {\bibinfo {volume} {7}},\ \bibinfo {pages} {041069}
  (\bibinfo {year} {2017})}\BibitemShut {NoStop}%
\bibitem [{\citenamefont {Song}\ \emph {et~al.}(2018)\citenamefont {Song},
  \citenamefont {Zhang}, \citenamefont {Fang},\ and\ \citenamefont
  {Fang}}]{song2018quantitative}%
  \BibitemOpen
  \bibfield  {author} {\bibinfo {author} {\bibfnamefont {Z.}~\bibnamefont
  {Song}}, \bibinfo {author} {\bibfnamefont {T.}~\bibnamefont {Zhang}},
  \bibinfo {author} {\bibfnamefont {Z.}~\bibnamefont {Fang}}, \ and\ \bibinfo
  {author} {\bibfnamefont {C.}~\bibnamefont {Fang}},\ }\href@noop {} {\bibfield
   {journal} {\bibinfo  {journal} {Nature communications}\ }\textbf {\bibinfo
  {volume} {9}},\ \bibinfo {pages} {1} (\bibinfo {year} {2018})}\BibitemShut
  {NoStop}%
\bibitem [{\citenamefont {Khalaf}\ \emph {et~al.}(2018)\citenamefont {Khalaf},
  \citenamefont {Po}, \citenamefont {Vishwanath},\ and\ \citenamefont
  {Watanabe}}]{PhysRevX.8.031070}%
  \BibitemOpen
  \bibfield  {author} {\bibinfo {author} {\bibfnamefont {E.}~\bibnamefont
  {Khalaf}}, \bibinfo {author} {\bibfnamefont {H.~C.}\ \bibnamefont {Po}},
  \bibinfo {author} {\bibfnamefont {A.}~\bibnamefont {Vishwanath}}, \ and\
  \bibinfo {author} {\bibfnamefont {H.}~\bibnamefont {Watanabe}},\ }\href
  {\doibase 10.1103/PhysRevX.8.031070} {\bibfield  {journal} {\bibinfo
  {journal} {Phys. Rev. X}\ }\textbf {\bibinfo {volume} {8}},\ \bibinfo {pages}
  {031070} (\bibinfo {year} {2018})}\BibitemShut {NoStop}%
\bibitem [{\citenamefont {Zhang}\ \emph
  {et~al.}(2019{\natexlab{a}})\citenamefont {Zhang}, \citenamefont {Jiang},
  \citenamefont {Song}, \citenamefont {Huang}, \citenamefont {He},
  \citenamefont {Fang}, \citenamefont {Weng},\ and\ \citenamefont
  {Fang}}]{zhang2019catalogue}%
  \BibitemOpen
  \bibfield  {author} {\bibinfo {author} {\bibfnamefont {T.}~\bibnamefont
  {Zhang}}, \bibinfo {author} {\bibfnamefont {Y.}~\bibnamefont {Jiang}},
  \bibinfo {author} {\bibfnamefont {Z.}~\bibnamefont {Song}}, \bibinfo {author}
  {\bibfnamefont {H.}~\bibnamefont {Huang}}, \bibinfo {author} {\bibfnamefont
  {Y.}~\bibnamefont {He}}, \bibinfo {author} {\bibfnamefont {Z.}~\bibnamefont
  {Fang}}, \bibinfo {author} {\bibfnamefont {H.}~\bibnamefont {Weng}}, \ and\
  \bibinfo {author} {\bibfnamefont {C.}~\bibnamefont {Fang}},\ }\href@noop {}
  {\bibfield  {journal} {\bibinfo  {journal} {Nature}\ }\textbf {\bibinfo
  {volume} {566}},\ \bibinfo {pages} {475} (\bibinfo {year}
  {2019}{\natexlab{a}})}\BibitemShut {NoStop}%
\bibitem [{\citenamefont {Tang}\ \emph {et~al.}(2019)\citenamefont {Tang},
  \citenamefont {Po}, \citenamefont {Vishwanath},\ and\ \citenamefont
  {Wan}}]{tang2019efficient}%
  \BibitemOpen
  \bibfield  {author} {\bibinfo {author} {\bibfnamefont {F.}~\bibnamefont
  {Tang}}, \bibinfo {author} {\bibfnamefont {H.~C.}\ \bibnamefont {Po}},
  \bibinfo {author} {\bibfnamefont {A.}~\bibnamefont {Vishwanath}}, \ and\
  \bibinfo {author} {\bibfnamefont {X.}~\bibnamefont {Wan}},\ }\href@noop {}
  {\bibfield  {journal} {\bibinfo  {journal} {Nature Physics}\ }\textbf
  {\bibinfo {volume} {15}},\ \bibinfo {pages} {470} (\bibinfo {year}
  {2019})}\BibitemShut {NoStop}%
\bibitem [{\citenamefont {Ueno}\ \emph {et~al.}(2013)\citenamefont {Ueno},
  \citenamefont {Yamakage}, \citenamefont {Tanaka},\ and\ \citenamefont
  {Sato}}]{PhysRevLett.111.087002}%
  \BibitemOpen
  \bibfield  {author} {\bibinfo {author} {\bibfnamefont {Y.}~\bibnamefont
  {Ueno}}, \bibinfo {author} {\bibfnamefont {A.}~\bibnamefont {Yamakage}},
  \bibinfo {author} {\bibfnamefont {Y.}~\bibnamefont {Tanaka}}, \ and\ \bibinfo
  {author} {\bibfnamefont {M.}~\bibnamefont {Sato}},\ }\href {\doibase
  10.1103/PhysRevLett.111.087002} {\bibfield  {journal} {\bibinfo  {journal}
  {Phys. Rev. Lett.}\ }\textbf {\bibinfo {volume} {111}},\ \bibinfo {pages}
  {087002} (\bibinfo {year} {2013})}\BibitemShut {NoStop}%
\bibitem [{\citenamefont {Tsutsumi}\ \emph {et~al.}(2013)\citenamefont
  {Tsutsumi}, \citenamefont {Ishikawa}, \citenamefont {Kawakami}, \citenamefont
  {Mizushima}, \citenamefont {Sato}, \citenamefont {Ichioka},\ and\
  \citenamefont {Machida}}]{doi:10.7566/JPSJ.82.113707}%
  \BibitemOpen
  \bibfield  {author} {\bibinfo {author} {\bibfnamefont {Y.}~\bibnamefont
  {Tsutsumi}}, \bibinfo {author} {\bibfnamefont {M.}~\bibnamefont {Ishikawa}},
  \bibinfo {author} {\bibfnamefont {T.}~\bibnamefont {Kawakami}}, \bibinfo
  {author} {\bibfnamefont {T.}~\bibnamefont {Mizushima}}, \bibinfo {author}
  {\bibfnamefont {M.}~\bibnamefont {Sato}}, \bibinfo {author} {\bibfnamefont
  {M.}~\bibnamefont {Ichioka}}, \ and\ \bibinfo {author} {\bibfnamefont
  {K.}~\bibnamefont {Machida}},\ }\href {\doibase 10.7566/JPSJ.82.113707}
  {\bibfield  {journal} {\bibinfo  {journal} {Journal of the Physical Society
  of Japan}\ }\textbf {\bibinfo {volume} {82}},\ \bibinfo {pages} {113707}
  (\bibinfo {year} {2013})},\ \Eprint
  {http://arxiv.org/abs/https://doi.org/10.7566/JPSJ.82.113707}
  {https://doi.org/10.7566/JPSJ.82.113707} \BibitemShut {NoStop}%
\bibitem [{\citenamefont {Zhang}\ \emph
  {et~al.}(2013{\natexlab{b}})\citenamefont {Zhang}, \citenamefont {Kane},\
  and\ \citenamefont {Mele}}]{PhysRevLett.111.056403}%
  \BibitemOpen
  \bibfield  {author} {\bibinfo {author} {\bibfnamefont {F.}~\bibnamefont
  {Zhang}}, \bibinfo {author} {\bibfnamefont {C.~L.}\ \bibnamefont {Kane}}, \
  and\ \bibinfo {author} {\bibfnamefont {E.~J.}\ \bibnamefont {Mele}},\ }\href
  {\doibase 10.1103/PhysRevLett.111.056403} {\bibfield  {journal} {\bibinfo
  {journal} {Phys. Rev. Lett.}\ }\textbf {\bibinfo {volume} {111}},\ \bibinfo
  {pages} {056403} (\bibinfo {year} {2013}{\natexlab{b}})}\BibitemShut
  {NoStop}%
\bibitem [{\citenamefont {Fang}\ \emph {et~al.}(2014)\citenamefont {Fang},
  \citenamefont {Gilbert},\ and\ \citenamefont
  {Bernevig}}]{PhysRevLett.112.106401}%
  \BibitemOpen
  \bibfield  {author} {\bibinfo {author} {\bibfnamefont {C.}~\bibnamefont
  {Fang}}, \bibinfo {author} {\bibfnamefont {M.~J.}\ \bibnamefont {Gilbert}}, \
  and\ \bibinfo {author} {\bibfnamefont {B.~A.}\ \bibnamefont {Bernevig}},\
  }\href {\doibase 10.1103/PhysRevLett.112.106401} {\bibfield  {journal}
  {\bibinfo  {journal} {Phys. Rev. Lett.}\ }\textbf {\bibinfo {volume} {112}},\
  \bibinfo {pages} {106401} (\bibinfo {year} {2014})}\BibitemShut {NoStop}%
\bibitem [{\citenamefont {Kobayashi}\ and\ \citenamefont
  {Sato}(2015)}]{rotation_Sato}%
  \BibitemOpen
  \bibfield  {author} {\bibinfo {author} {\bibfnamefont {S.}~\bibnamefont
  {Kobayashi}}\ and\ \bibinfo {author} {\bibfnamefont {M.}~\bibnamefont
  {Sato}},\ }\href {\doibase 10.1103/PhysRevLett.115.187001} {\bibfield
  {journal} {\bibinfo  {journal} {Phys. Rev. Lett.}\ }\textbf {\bibinfo
  {volume} {115}},\ \bibinfo {pages} {187001} (\bibinfo {year}
  {2015})}\BibitemShut {NoStop}%
\bibitem [{\citenamefont {Fang}\ \emph {et~al.}(2017)\citenamefont {Fang},
  \citenamefont {Bernevig},\ and\ \citenamefont {Gilbert}}]{rotation_CFang}%
  \BibitemOpen
  \bibfield  {author} {\bibinfo {author} {\bibfnamefont {C.}~\bibnamefont
  {Fang}}, \bibinfo {author} {\bibfnamefont {B.~A.}\ \bibnamefont {Bernevig}},
  \ and\ \bibinfo {author} {\bibfnamefont {M.~J.}\ \bibnamefont {Gilbert}},\
  }\href@noop {} {\bibfield  {journal} {\bibinfo  {journal} {arXiv preprint
  arXiv:1701.01944}\ } (\bibinfo {year} {2017})}\BibitemShut {NoStop}%
\bibitem [{\citenamefont {Zhang}\ and\ \citenamefont
  {Liu}(2018)}]{rotation_RXZhang}%
  \BibitemOpen
  \bibfield  {author} {\bibinfo {author} {\bibfnamefont {R.-X.}\ \bibnamefont
  {Zhang}}\ and\ \bibinfo {author} {\bibfnamefont {C.-X.}\ \bibnamefont
  {Liu}},\ }\href {\doibase 10.1103/PhysRevLett.120.156802} {\bibfield
  {journal} {\bibinfo  {journal} {Phys. Rev. Lett.}\ }\textbf {\bibinfo
  {volume} {120}},\ \bibinfo {pages} {156802} (\bibinfo {year}
  {2018})}\BibitemShut {NoStop}%
\bibitem [{\citenamefont {Qin}\ \emph {et~al.}(2019{\natexlab{a}})\citenamefont
  {Qin}, \citenamefont {Hu}, \citenamefont {Le}, \citenamefont {Zeng},
  \citenamefont {Zhang}, \citenamefont {Fang},\ and\ \citenamefont
  {Hu}}]{PhysRevLett.123.027003}%
  \BibitemOpen
  \bibfield  {author} {\bibinfo {author} {\bibfnamefont {S.}~\bibnamefont
  {Qin}}, \bibinfo {author} {\bibfnamefont {L.}~\bibnamefont {Hu}}, \bibinfo
  {author} {\bibfnamefont {C.}~\bibnamefont {Le}}, \bibinfo {author}
  {\bibfnamefont {J.}~\bibnamefont {Zeng}}, \bibinfo {author} {\bibfnamefont
  {F.-c.}\ \bibnamefont {Zhang}}, \bibinfo {author} {\bibfnamefont
  {C.}~\bibnamefont {Fang}}, \ and\ \bibinfo {author} {\bibfnamefont
  {J.}~\bibnamefont {Hu}},\ }\href {\doibase 10.1103/PhysRevLett.123.027003}
  {\bibfield  {journal} {\bibinfo  {journal} {Phys. Rev. Lett.}\ }\textbf
  {\bibinfo {volume} {123}},\ \bibinfo {pages} {027003} (\bibinfo {year}
  {2019}{\natexlab{a}})}\BibitemShut {NoStop}%
\bibitem [{\citenamefont {K\"onig}\ and\ \citenamefont
  {Coleman}(2019)}]{PhysRevLett.122.207001}%
  \BibitemOpen
  \bibfield  {author} {\bibinfo {author} {\bibfnamefont {E.~J.}\ \bibnamefont
  {K\"onig}}\ and\ \bibinfo {author} {\bibfnamefont {P.}~\bibnamefont
  {Coleman}},\ }\href {\doibase 10.1103/PhysRevLett.122.207001} {\bibfield
  {journal} {\bibinfo  {journal} {Phys. Rev. Lett.}\ }\textbf {\bibinfo
  {volume} {122}},\ \bibinfo {pages} {207001} (\bibinfo {year}
  {2019})}\BibitemShut {NoStop}%
\bibitem [{\citenamefont {Qin}\ \emph {et~al.}(2019{\natexlab{b}})\citenamefont
  {Qin}, \citenamefont {Hu}, \citenamefont {Wu}, \citenamefont {Dai},
  \citenamefont {Fang}, \citenamefont {Zhang},\ and\ \citenamefont
  {Hu}}]{qin2019topological}%
  \BibitemOpen
  \bibfield  {author} {\bibinfo {author} {\bibfnamefont {S.}~\bibnamefont
  {Qin}}, \bibinfo {author} {\bibfnamefont {L.}~\bibnamefont {Hu}}, \bibinfo
  {author} {\bibfnamefont {X.}~\bibnamefont {Wu}}, \bibinfo {author}
  {\bibfnamefont {X.}~\bibnamefont {Dai}}, \bibinfo {author} {\bibfnamefont
  {C.}~\bibnamefont {Fang}}, \bibinfo {author} {\bibfnamefont {F.-C.}\
  \bibnamefont {Zhang}}, \ and\ \bibinfo {author} {\bibfnamefont
  {J.}~\bibnamefont {Hu}},\ }\href@noop {} {\bibfield  {journal} {\bibinfo
  {journal} {Science Bulletin}\ }\textbf {\bibinfo {volume} {64}},\ \bibinfo
  {pages} {1207} (\bibinfo {year} {2019}{\natexlab{b}})}\BibitemShut {NoStop}%
\bibitem [{\citenamefont {Zhang}\ \emph
  {et~al.}(2020{\natexlab{a}})\citenamefont {Zhang}, \citenamefont {Hsu},\ and\
  \citenamefont {Das~Sarma}}]{rotation_HOTDS}%
  \BibitemOpen
  \bibfield  {author} {\bibinfo {author} {\bibfnamefont {R.-X.}\ \bibnamefont
  {Zhang}}, \bibinfo {author} {\bibfnamefont {Y.-T.}\ \bibnamefont {Hsu}}, \
  and\ \bibinfo {author} {\bibfnamefont {S.}~\bibnamefont {Das~Sarma}},\ }\href
  {\doibase 10.1103/PhysRevB.102.094503} {\bibfield  {journal} {\bibinfo
  {journal} {Phys. Rev. B}\ }\textbf {\bibinfo {volume} {102}},\ \bibinfo
  {pages} {094503} (\bibinfo {year} {2020}{\natexlab{a}})}\BibitemShut
  {NoStop}%
\bibitem [{\citenamefont {Wang}\ and\ \citenamefont
  {Liu}(2016)}]{PhysRevB.93.020505}%
  \BibitemOpen
  \bibfield  {author} {\bibinfo {author} {\bibfnamefont {Q.-Z.}\ \bibnamefont
  {Wang}}\ and\ \bibinfo {author} {\bibfnamefont {C.-X.}\ \bibnamefont {Liu}},\
  }\href {\doibase 10.1103/PhysRevB.93.020505} {\bibfield  {journal} {\bibinfo
  {journal} {Phys. Rev. B}\ }\textbf {\bibinfo {volume} {93}},\ \bibinfo
  {pages} {020505} (\bibinfo {year} {2016})}\BibitemShut {NoStop}%
\bibitem [{\citenamefont {Shiozaki}\ \emph {et~al.}(2016)\citenamefont
  {Shiozaki}, \citenamefont {Sato},\ and\ \citenamefont
  {Gomi}}]{PhysRevB.93.195413}%
  \BibitemOpen
  \bibfield  {author} {\bibinfo {author} {\bibfnamefont {K.}~\bibnamefont
  {Shiozaki}}, \bibinfo {author} {\bibfnamefont {M.}~\bibnamefont {Sato}}, \
  and\ \bibinfo {author} {\bibfnamefont {K.}~\bibnamefont {Gomi}},\ }\href
  {\doibase 10.1103/PhysRevB.93.195413} {\bibfield  {journal} {\bibinfo
  {journal} {Phys. Rev. B}\ }\textbf {\bibinfo {volume} {93}},\ \bibinfo
  {pages} {195413} (\bibinfo {year} {2016})}\BibitemShut {NoStop}%
\bibitem [{\citenamefont {Daido}\ \emph {et~al.}(2019)\citenamefont {Daido},
  \citenamefont {Yoshida},\ and\ \citenamefont
  {Yanase}}]{PhysRevLett.122.227001}%
  \BibitemOpen
  \bibfield  {author} {\bibinfo {author} {\bibfnamefont {A.}~\bibnamefont
  {Daido}}, \bibinfo {author} {\bibfnamefont {T.}~\bibnamefont {Yoshida}}, \
  and\ \bibinfo {author} {\bibfnamefont {Y.}~\bibnamefont {Yanase}},\ }\href
  {\doibase 10.1103/PhysRevLett.122.227001} {\bibfield  {journal} {\bibinfo
  {journal} {Phys. Rev. Lett.}\ }\textbf {\bibinfo {volume} {122}},\ \bibinfo
  {pages} {227001} (\bibinfo {year} {2019})}\BibitemShut {NoStop}%
\bibitem [{\citenamefont {Schnyder}\ \emph {et~al.}(2008)\citenamefont
  {Schnyder}, \citenamefont {Ryu}, \citenamefont {Furusaki},\ and\
  \citenamefont {Ludwig}}]{classification1}%
  \BibitemOpen
  \bibfield  {author} {\bibinfo {author} {\bibfnamefont {A.~P.}\ \bibnamefont
  {Schnyder}}, \bibinfo {author} {\bibfnamefont {S.}~\bibnamefont {Ryu}},
  \bibinfo {author} {\bibfnamefont {A.}~\bibnamefont {Furusaki}}, \ and\
  \bibinfo {author} {\bibfnamefont {A.~W.~W.}\ \bibnamefont {Ludwig}},\ }\href
  {\doibase 10.1103/PhysRevB.78.195125} {\bibfield  {journal} {\bibinfo
  {journal} {Phys. Rev. B}\ }\textbf {\bibinfo {volume} {78}},\ \bibinfo
  {pages} {195125} (\bibinfo {year} {2008})}\BibitemShut {NoStop}%
\bibitem [{\citenamefont {Ryu}\ \emph {et~al.}(2010)\citenamefont {Ryu},
  \citenamefont {Schnyder}, \citenamefont {Furusaki},\ and\ \citenamefont
  {Ludwig}}]{classification2}%
  \BibitemOpen
  \bibfield  {author} {\bibinfo {author} {\bibfnamefont {S.}~\bibnamefont
  {Ryu}}, \bibinfo {author} {\bibfnamefont {A.~P.}\ \bibnamefont {Schnyder}},
  \bibinfo {author} {\bibfnamefont {A.}~\bibnamefont {Furusaki}}, \ and\
  \bibinfo {author} {\bibfnamefont {A.~W.~W.}\ \bibnamefont {Ludwig}},\ }\href
  {\doibase 10.1088/1367-2630/12/6/065010} {\bibfield  {journal} {\bibinfo
  {journal} {New Journal of Physics}\ }\textbf {\bibinfo {volume} {12}},\
  \bibinfo {pages} {065010} (\bibinfo {year} {2010})}\BibitemShut {NoStop}%
\bibitem [{\citenamefont {Mazin}(2011)}]{PhysRevB.84.024529}%
  \BibitemOpen
  \bibfield  {author} {\bibinfo {author} {\bibfnamefont {I.~I.}\ \bibnamefont
  {Mazin}},\ }\href {\doibase 10.1103/PhysRevB.84.024529} {\bibfield  {journal}
  {\bibinfo  {journal} {Phys. Rev. B}\ }\textbf {\bibinfo {volume} {84}},\
  \bibinfo {pages} {024529} (\bibinfo {year} {2011})}\BibitemShut {NoStop}%
\bibitem [{\citenamefont {Hirschfeld}\ \emph {et~al.}(2011)\citenamefont
  {Hirschfeld}, \citenamefont {Korshunov},\ and\ \citenamefont
  {Mazin}}]{Hirschfeld_2011}%
  \BibitemOpen
  \bibfield  {author} {\bibinfo {author} {\bibfnamefont {P.~J.}\ \bibnamefont
  {Hirschfeld}}, \bibinfo {author} {\bibfnamefont {M.~M.}\ \bibnamefont
  {Korshunov}}, \ and\ \bibinfo {author} {\bibfnamefont {I.~I.}\ \bibnamefont
  {Mazin}},\ }\href {\doibase 10.1088/0034-4885/74/12/124508} {\bibfield
  {journal} {\bibinfo  {journal} {Reports on Progress in Physics}\ }\textbf
  {\bibinfo {volume} {74}},\ \bibinfo {pages} {124508} (\bibinfo {year}
  {2011})}\BibitemShut {NoStop}%
\bibitem [{\citenamefont {Hu}(2013)}]{PhysRevX.3.031004}%
  \BibitemOpen
  \bibfield  {author} {\bibinfo {author} {\bibfnamefont {J.}~\bibnamefont
  {Hu}},\ }\href {\doibase 10.1103/PhysRevX.3.031004} {\bibfield  {journal}
  {\bibinfo  {journal} {Phys. Rev. X}\ }\textbf {\bibinfo {volume} {3}},\
  \bibinfo {pages} {031004} (\bibinfo {year} {2013})}\BibitemShut {NoStop}%
\bibitem [{\citenamefont {Zhang}\ \emph {et~al.}(2014)\citenamefont {Zhang},
  \citenamefont {Liu}, \citenamefont {Luo}, \citenamefont {Freeman},\ and\
  \citenamefont {Zunger}}]{zhang2014hidden}%
  \BibitemOpen
  \bibfield  {author} {\bibinfo {author} {\bibfnamefont {X.}~\bibnamefont
  {Zhang}}, \bibinfo {author} {\bibfnamefont {Q.}~\bibnamefont {Liu}}, \bibinfo
  {author} {\bibfnamefont {J.-W.}\ \bibnamefont {Luo}}, \bibinfo {author}
  {\bibfnamefont {A.~J.}\ \bibnamefont {Freeman}}, \ and\ \bibinfo {author}
  {\bibfnamefont {A.}~\bibnamefont {Zunger}},\ }\href@noop {} {\bibfield
  {journal} {\bibinfo  {journal} {Nature Physics}\ }\textbf {\bibinfo {volume}
  {10}},\ \bibinfo {pages} {387} (\bibinfo {year} {2014})}\BibitemShut
  {NoStop}%
\bibitem [{\citenamefont {Wu}\ \emph {et~al.}(2017)\citenamefont {Wu},
  \citenamefont {Sumida}, \citenamefont {Miyamoto}, \citenamefont {Taguchi},
  \citenamefont {Yoshikawa}, \citenamefont {Kimura}, \citenamefont {Ueda},
  \citenamefont {Arita}, \citenamefont {Nagao}, \citenamefont {Watauchi} \emph
  {et~al.}}]{wu2017direct}%
  \BibitemOpen
  \bibfield  {author} {\bibinfo {author} {\bibfnamefont {S.-L.}\ \bibnamefont
  {Wu}}, \bibinfo {author} {\bibfnamefont {K.}~\bibnamefont {Sumida}}, \bibinfo
  {author} {\bibfnamefont {K.}~\bibnamefont {Miyamoto}}, \bibinfo {author}
  {\bibfnamefont {K.}~\bibnamefont {Taguchi}}, \bibinfo {author} {\bibfnamefont
  {T.}~\bibnamefont {Yoshikawa}}, \bibinfo {author} {\bibfnamefont
  {A.}~\bibnamefont {Kimura}}, \bibinfo {author} {\bibfnamefont
  {Y.}~\bibnamefont {Ueda}}, \bibinfo {author} {\bibfnamefont {M.}~\bibnamefont
  {Arita}}, \bibinfo {author} {\bibfnamefont {M.}~\bibnamefont {Nagao}},
  \bibinfo {author} {\bibfnamefont {S.}~\bibnamefont {Watauchi}},  \emph
  {et~al.},\ }\href@noop {} {\bibfield  {journal} {\bibinfo  {journal} {Nature
  communications}\ }\textbf {\bibinfo {volume} {8}},\ \bibinfo {pages} {1}
  (\bibinfo {year} {2017})}\BibitemShut {NoStop}%
\bibitem [{\citenamefont {Zhang}\ \emph
  {et~al.}(2020{\natexlab{b}})\citenamefont {Zhang}, \citenamefont {Liu},
  \citenamefont {Sun}, \citenamefont {Zhao}, \citenamefont {Xu},\ and\
  \citenamefont {Liu}}]{Zhang_2020}%
  \BibitemOpen
  \bibfield  {author} {\bibinfo {author} {\bibfnamefont {Y.}~\bibnamefont
  {Zhang}}, \bibinfo {author} {\bibfnamefont {P.}~\bibnamefont {Liu}}, \bibinfo
  {author} {\bibfnamefont {H.}~\bibnamefont {Sun}}, \bibinfo {author}
  {\bibfnamefont {S.}~\bibnamefont {Zhao}}, \bibinfo {author} {\bibfnamefont
  {H.}~\bibnamefont {Xu}}, \ and\ \bibinfo {author} {\bibfnamefont
  {Q.}~\bibnamefont {Liu}},\ }\href {\doibase 10.1088/0256-307x/37/8/087105}
  {\bibfield  {journal} {\bibinfo  {journal} {Chinese Physics Letters}\
  }\textbf {\bibinfo {volume} {37}},\ \bibinfo {pages} {087105} (\bibinfo
  {year} {2020}{\natexlab{b}})}\BibitemShut {NoStop}%
\bibitem [{foo()}]{footnote1}%
  \BibitemOpen
  \href@noop {} {\bibinfo  {journal} {The commutation/anticommutation relation
  in Eq.(5) is true only in the sense of the result in Eq.(4). In general the
  commutation relation between a unitary operator and an antiunitary operator
  is not well defined, considering that the relation can be changed by a U(1)
  gauge}\ }\BibitemShut {NoStop}%
\bibitem [{\citenamefont {Sato}\ \emph {et~al.}(2011)\citenamefont {Sato},
  \citenamefont {Tanaka}, \citenamefont {Yada},\ and\ \citenamefont
  {Yokoyama}}]{PhysRevB.83.224511}%
  \BibitemOpen
\bibfield  {journal} {  }\bibfield  {author} {\bibinfo {author} {\bibfnamefont
  {M.}~\bibnamefont {Sato}}, \bibinfo {author} {\bibfnamefont {Y.}~\bibnamefont
  {Tanaka}}, \bibinfo {author} {\bibfnamefont {K.}~\bibnamefont {Yada}}, \ and\
  \bibinfo {author} {\bibfnamefont {T.}~\bibnamefont {Yokoyama}},\ }\href
  {\doibase 10.1103/PhysRevB.83.224511} {\bibfield  {journal} {\bibinfo
  {journal} {Phys. Rev. B}\ }\textbf {\bibinfo {volume} {83}},\ \bibinfo
  {pages} {224511} (\bibinfo {year} {2011})}\BibitemShut {NoStop}%
\bibitem [{\citenamefont {Wu}\ \emph {et~al.}(2020{\natexlab{b}})\citenamefont
  {Wu}, \citenamefont {Benalcazar}, \citenamefont {Li}, \citenamefont
  {Thomale}, \citenamefont {Liu},\ and\ \citenamefont
  {Hu}}]{PhysRevX.10.041014}%
  \BibitemOpen
  \bibfield  {author} {\bibinfo {author} {\bibfnamefont {X.}~\bibnamefont
  {Wu}}, \bibinfo {author} {\bibfnamefont {W.~A.}\ \bibnamefont {Benalcazar}},
  \bibinfo {author} {\bibfnamefont {Y.}~\bibnamefont {Li}}, \bibinfo {author}
  {\bibfnamefont {R.}~\bibnamefont {Thomale}}, \bibinfo {author} {\bibfnamefont
  {C.-X.}\ \bibnamefont {Liu}}, \ and\ \bibinfo {author} {\bibfnamefont
  {J.}~\bibnamefont {Hu}},\ }\href {\doibase 10.1103/PhysRevX.10.041014}
  {\bibfield  {journal} {\bibinfo  {journal} {Phys. Rev. X}\ }\textbf {\bibinfo
  {volume} {10}},\ \bibinfo {pages} {041014} (\bibinfo {year}
  {2020}{\natexlab{b}})}\BibitemShut {NoStop}%
\bibitem [{\citenamefont {Benalcazar}\ \emph
  {et~al.}(2017{\natexlab{a}})\citenamefont {Benalcazar}, \citenamefont
  {Bernevig},\ and\ \citenamefont {Hughes}}]{benalcazar2017quantized}%
  \BibitemOpen
  \bibfield  {author} {\bibinfo {author} {\bibfnamefont {W.~A.}\ \bibnamefont
  {Benalcazar}}, \bibinfo {author} {\bibfnamefont {B.~A.}\ \bibnamefont
  {Bernevig}}, \ and\ \bibinfo {author} {\bibfnamefont {T.~L.}\ \bibnamefont
  {Hughes}},\ }\href@noop {} {\bibfield  {journal} {\bibinfo  {journal}
  {Science}\ }\textbf {\bibinfo {volume} {357}},\ \bibinfo {pages} {61}
  (\bibinfo {year} {2017}{\natexlab{a}})}\BibitemShut {NoStop}%
\bibitem [{\citenamefont {Benalcazar}\ \emph
  {et~al.}(2017{\natexlab{b}})\citenamefont {Benalcazar}, \citenamefont
  {Bernevig},\ and\ \citenamefont {Hughes}}]{PhysRevB.96.245115}%
  \BibitemOpen
  \bibfield  {author} {\bibinfo {author} {\bibfnamefont {W.~A.}\ \bibnamefont
  {Benalcazar}}, \bibinfo {author} {\bibfnamefont {B.~A.}\ \bibnamefont
  {Bernevig}}, \ and\ \bibinfo {author} {\bibfnamefont {T.~L.}\ \bibnamefont
  {Hughes}},\ }\href {\doibase 10.1103/PhysRevB.96.245115} {\bibfield
  {journal} {\bibinfo  {journal} {Phys. Rev. B}\ }\textbf {\bibinfo {volume}
  {96}},\ \bibinfo {pages} {245115} (\bibinfo {year}
  {2017}{\natexlab{b}})}\BibitemShut {NoStop}%
\bibitem [{\citenamefont {Song}\ \emph {et~al.}(2017)\citenamefont {Song},
  \citenamefont {Fang},\ and\ \citenamefont {Fang}}]{PhysRevLett.119.246402}%
  \BibitemOpen
  \bibfield  {author} {\bibinfo {author} {\bibfnamefont {Z.}~\bibnamefont
  {Song}}, \bibinfo {author} {\bibfnamefont {Z.}~\bibnamefont {Fang}}, \ and\
  \bibinfo {author} {\bibfnamefont {C.}~\bibnamefont {Fang}},\ }\href {\doibase
  10.1103/PhysRevLett.119.246402} {\bibfield  {journal} {\bibinfo  {journal}
  {Phys. Rev. Lett.}\ }\textbf {\bibinfo {volume} {119}},\ \bibinfo {pages}
  {246402} (\bibinfo {year} {2017})}\BibitemShut {NoStop}%
\bibitem [{\citenamefont {Schindler}\ \emph {et~al.}(2018)\citenamefont
  {Schindler}, \citenamefont {Cook}, \citenamefont {Vergniory}, \citenamefont
  {Wang}, \citenamefont {Parkin}, \citenamefont {Bernevig},\ and\ \citenamefont
  {Neupert}}]{schindler2018higher}%
  \BibitemOpen
  \bibfield  {author} {\bibinfo {author} {\bibfnamefont {F.}~\bibnamefont
  {Schindler}}, \bibinfo {author} {\bibfnamefont {A.~M.}\ \bibnamefont {Cook}},
  \bibinfo {author} {\bibfnamefont {M.~G.}\ \bibnamefont {Vergniory}}, \bibinfo
  {author} {\bibfnamefont {Z.}~\bibnamefont {Wang}}, \bibinfo {author}
  {\bibfnamefont {S.~S.}\ \bibnamefont {Parkin}}, \bibinfo {author}
  {\bibfnamefont {B.~A.}\ \bibnamefont {Bernevig}}, \ and\ \bibinfo {author}
  {\bibfnamefont {T.}~\bibnamefont {Neupert}},\ }\href@noop {} {\bibfield
  {journal} {\bibinfo  {journal} {Science advances}\ }\textbf {\bibinfo
  {volume} {4}},\ \bibinfo {pages} {eaat0346} (\bibinfo {year}
  {2018})}\BibitemShut {NoStop}%
\bibitem [{\citenamefont {Langbehn}\ \emph {et~al.}(2017)\citenamefont
  {Langbehn}, \citenamefont {Peng}, \citenamefont {Trifunovic}, \citenamefont
  {von Oppen},\ and\ \citenamefont {Brouwer}}]{PhysRevLett.119.246401}%
  \BibitemOpen
  \bibfield  {author} {\bibinfo {author} {\bibfnamefont {J.}~\bibnamefont
  {Langbehn}}, \bibinfo {author} {\bibfnamefont {Y.}~\bibnamefont {Peng}},
  \bibinfo {author} {\bibfnamefont {L.}~\bibnamefont {Trifunovic}}, \bibinfo
  {author} {\bibfnamefont {F.}~\bibnamefont {von Oppen}}, \ and\ \bibinfo
  {author} {\bibfnamefont {P.~W.}\ \bibnamefont {Brouwer}},\ }\href {\doibase
  10.1103/PhysRevLett.119.246401} {\bibfield  {journal} {\bibinfo  {journal}
  {Phys. Rev. Lett.}\ }\textbf {\bibinfo {volume} {119}},\ \bibinfo {pages}
  {246401} (\bibinfo {year} {2017})}\BibitemShut {NoStop}%
\bibitem [{\citenamefont {Jackiw}\ and\ \citenamefont
  {Rebbi}(1976)}]{PhysRevD.13.3398}%
  \BibitemOpen
  \bibfield  {author} {\bibinfo {author} {\bibfnamefont {R.}~\bibnamefont
  {Jackiw}}\ and\ \bibinfo {author} {\bibfnamefont {C.}~\bibnamefont {Rebbi}},\
  }\href {\doibase 10.1103/PhysRevD.13.3398} {\bibfield  {journal} {\bibinfo
  {journal} {Phys. Rev. D}\ }\textbf {\bibinfo {volume} {13}},\ \bibinfo
  {pages} {3398} (\bibinfo {year} {1976})}\BibitemShut {NoStop}%
\bibitem [{\citenamefont {Yan}\ \emph {et~al.}(2018)\citenamefont {Yan},
  \citenamefont {Song},\ and\ \citenamefont {Wang}}]{PhysRevLett.121.096803}%
  \BibitemOpen
  \bibfield  {author} {\bibinfo {author} {\bibfnamefont {Z.}~\bibnamefont
  {Yan}}, \bibinfo {author} {\bibfnamefont {F.}~\bibnamefont {Song}}, \ and\
  \bibinfo {author} {\bibfnamefont {Z.}~\bibnamefont {Wang}},\ }\href {\doibase
  10.1103/PhysRevLett.121.096803} {\bibfield  {journal} {\bibinfo  {journal}
  {Phys. Rev. Lett.}\ }\textbf {\bibinfo {volume} {121}},\ \bibinfo {pages}
  {096803} (\bibinfo {year} {2018})}\BibitemShut {NoStop}%
\bibitem [{\citenamefont {Wang}\ \emph
  {et~al.}(2018{\natexlab{b}})\citenamefont {Wang}, \citenamefont {Liu},
  \citenamefont {Lu},\ and\ \citenamefont {Zhang}}]{PhysRevLett.121.186801}%
  \BibitemOpen
  \bibfield  {author} {\bibinfo {author} {\bibfnamefont {Q.}~\bibnamefont
  {Wang}}, \bibinfo {author} {\bibfnamefont {C.-C.}\ \bibnamefont {Liu}},
  \bibinfo {author} {\bibfnamefont {Y.-M.}\ \bibnamefont {Lu}}, \ and\ \bibinfo
  {author} {\bibfnamefont {F.}~\bibnamefont {Zhang}},\ }\href {\doibase
  10.1103/PhysRevLett.121.186801} {\bibfield  {journal} {\bibinfo  {journal}
  {Phys. Rev. Lett.}\ }\textbf {\bibinfo {volume} {121}},\ \bibinfo {pages}
  {186801} (\bibinfo {year} {2018}{\natexlab{b}})}\BibitemShut {NoStop}%
\bibitem [{\citenamefont {Zhang}\ \emph
  {et~al.}(2019{\natexlab{b}})\citenamefont {Zhang}, \citenamefont {Cole},\
  and\ \citenamefont {Das~Sarma}}]{PhysRevLett.122.187001}%
  \BibitemOpen
  \bibfield  {author} {\bibinfo {author} {\bibfnamefont {R.-X.}\ \bibnamefont
  {Zhang}}, \bibinfo {author} {\bibfnamefont {W.~S.}\ \bibnamefont {Cole}}, \
  and\ \bibinfo {author} {\bibfnamefont {S.}~\bibnamefont {Das~Sarma}},\ }\href
  {\doibase 10.1103/PhysRevLett.122.187001} {\bibfield  {journal} {\bibinfo
  {journal} {Phys. Rev. Lett.}\ }\textbf {\bibinfo {volume} {122}},\ \bibinfo
  {pages} {187001} (\bibinfo {year} {2019}{\natexlab{b}})}\BibitemShut
  {NoStop}%
\bibitem [{\citenamefont {Zhang}\ \emph {et~al.}(2011)\citenamefont {Zhang},
  \citenamefont {Yang}, \citenamefont {Xu}, \citenamefont {Ye}, \citenamefont
  {Chen}, \citenamefont {He}, \citenamefont {Xu}, \citenamefont {Jiang},
  \citenamefont {Xie}, \citenamefont {Ying} \emph
  {et~al.}}]{zhang2011nodeless}%
  \BibitemOpen
  \bibfield  {author} {\bibinfo {author} {\bibfnamefont {Y.}~\bibnamefont
  {Zhang}}, \bibinfo {author} {\bibfnamefont {L.}~\bibnamefont {Yang}},
  \bibinfo {author} {\bibfnamefont {M.}~\bibnamefont {Xu}}, \bibinfo {author}
  {\bibfnamefont {Z.}~\bibnamefont {Ye}}, \bibinfo {author} {\bibfnamefont
  {F.}~\bibnamefont {Chen}}, \bibinfo {author} {\bibfnamefont {C.}~\bibnamefont
  {He}}, \bibinfo {author} {\bibfnamefont {H.}~\bibnamefont {Xu}}, \bibinfo
  {author} {\bibfnamefont {J.}~\bibnamefont {Jiang}}, \bibinfo {author}
  {\bibfnamefont {B.}~\bibnamefont {Xie}}, \bibinfo {author} {\bibfnamefont
  {J.}~\bibnamefont {Ying}},  \emph {et~al.},\ }\href@noop {} {\bibfield
  {journal} {\bibinfo  {journal} {Nature materials}\ }\textbf {\bibinfo
  {volume} {10}},\ \bibinfo {pages} {273} (\bibinfo {year} {2011})}\BibitemShut
  {NoStop}%
\bibitem [{\citenamefont {Qian}\ \emph {et~al.}(2011)\citenamefont {Qian},
  \citenamefont {Wang}, \citenamefont {Jin}, \citenamefont {Zhang},
  \citenamefont {Richard}, \citenamefont {Xu}, \citenamefont {Dai},
  \citenamefont {Fang}, \citenamefont {Guo}, \citenamefont {Chen},\ and\
  \citenamefont {Ding}}]{PhysRevLett.106.187001}%
  \BibitemOpen
  \bibfield  {author} {\bibinfo {author} {\bibfnamefont {T.}~\bibnamefont
  {Qian}}, \bibinfo {author} {\bibfnamefont {X.-P.}\ \bibnamefont {Wang}},
  \bibinfo {author} {\bibfnamefont {W.-C.}\ \bibnamefont {Jin}}, \bibinfo
  {author} {\bibfnamefont {P.}~\bibnamefont {Zhang}}, \bibinfo {author}
  {\bibfnamefont {P.}~\bibnamefont {Richard}}, \bibinfo {author} {\bibfnamefont
  {G.}~\bibnamefont {Xu}}, \bibinfo {author} {\bibfnamefont {X.}~\bibnamefont
  {Dai}}, \bibinfo {author} {\bibfnamefont {Z.}~\bibnamefont {Fang}}, \bibinfo
  {author} {\bibfnamefont {J.-G.}\ \bibnamefont {Guo}}, \bibinfo {author}
  {\bibfnamefont {X.-L.}\ \bibnamefont {Chen}}, \ and\ \bibinfo {author}
  {\bibfnamefont {H.}~\bibnamefont {Ding}},\ }\href {\doibase
  10.1103/PhysRevLett.106.187001} {\bibfield  {journal} {\bibinfo  {journal}
  {Phys. Rev. Lett.}\ }\textbf {\bibinfo {volume} {106}},\ \bibinfo {pages}
  {187001} (\bibinfo {year} {2011})}\BibitemShut {NoStop}%
\bibitem [{\citenamefont {Liu}\ \emph {et~al.}(2012)\citenamefont {Liu},
  \citenamefont {Zhang}, \citenamefont {Mou}, \citenamefont {He}, \citenamefont
  {Ou}, \citenamefont {Wang}, \citenamefont {Li}, \citenamefont {Wang},
  \citenamefont {Zhao}, \citenamefont {He} \emph {et~al.}}]{liu2012electronic}%
  \BibitemOpen
  \bibfield  {author} {\bibinfo {author} {\bibfnamefont {D.}~\bibnamefont
  {Liu}}, \bibinfo {author} {\bibfnamefont {W.}~\bibnamefont {Zhang}}, \bibinfo
  {author} {\bibfnamefont {D.}~\bibnamefont {Mou}}, \bibinfo {author}
  {\bibfnamefont {J.}~\bibnamefont {He}}, \bibinfo {author} {\bibfnamefont
  {Y.-B.}\ \bibnamefont {Ou}}, \bibinfo {author} {\bibfnamefont {Q.-Y.}\
  \bibnamefont {Wang}}, \bibinfo {author} {\bibfnamefont {Z.}~\bibnamefont
  {Li}}, \bibinfo {author} {\bibfnamefont {L.}~\bibnamefont {Wang}}, \bibinfo
  {author} {\bibfnamefont {L.}~\bibnamefont {Zhao}}, \bibinfo {author}
  {\bibfnamefont {S.}~\bibnamefont {He}},  \emph {et~al.},\ }\href@noop {}
  {\bibfield  {journal} {\bibinfo  {journal} {Nature communications}\ }\textbf
  {\bibinfo {volume} {3}},\ \bibinfo {pages} {1} (\bibinfo {year}
  {2012})}\BibitemShut {NoStop}%
\bibitem [{\citenamefont {Zhao}\ \emph {et~al.}(2016)\citenamefont {Zhao},
  \citenamefont {Liang}, \citenamefont {Yuan}, \citenamefont {Hu},
  \citenamefont {Liu}, \citenamefont {Huang}, \citenamefont {He}, \citenamefont
  {Shen}, \citenamefont {Xu}, \citenamefont {Liu} \emph
  {et~al.}}]{zhao2016common}%
  \BibitemOpen
  \bibfield  {author} {\bibinfo {author} {\bibfnamefont {L.}~\bibnamefont
  {Zhao}}, \bibinfo {author} {\bibfnamefont {A.}~\bibnamefont {Liang}},
  \bibinfo {author} {\bibfnamefont {D.}~\bibnamefont {Yuan}}, \bibinfo {author}
  {\bibfnamefont {Y.}~\bibnamefont {Hu}}, \bibinfo {author} {\bibfnamefont
  {D.}~\bibnamefont {Liu}}, \bibinfo {author} {\bibfnamefont {J.}~\bibnamefont
  {Huang}}, \bibinfo {author} {\bibfnamefont {S.}~\bibnamefont {He}}, \bibinfo
  {author} {\bibfnamefont {B.}~\bibnamefont {Shen}}, \bibinfo {author}
  {\bibfnamefont {Y.}~\bibnamefont {Xu}}, \bibinfo {author} {\bibfnamefont
  {X.}~\bibnamefont {Liu}},  \emph {et~al.},\ }\href@noop {} {\bibfield
  {journal} {\bibinfo  {journal} {Nature communications}\ }\textbf {\bibinfo
  {volume} {7}},\ \bibinfo {pages} {1} (\bibinfo {year} {2016})}\BibitemShut
  {NoStop}%
\bibitem [{\citenamefont {Dai}(2015)}]{RevModPhys.87.855}%
  \BibitemOpen
  \bibfield  {author} {\bibinfo {author} {\bibfnamefont {P.}~\bibnamefont
  {Dai}},\ }\href {\doibase 10.1103/RevModPhys.87.855} {\bibfield  {journal}
  {\bibinfo  {journal} {Rev. Mod. Phys.}\ }\textbf {\bibinfo {volume} {87}},\
  \bibinfo {pages} {855} (\bibinfo {year} {2015})}\BibitemShut {NoStop}%
\bibitem [{\citenamefont {Huang}\ and\ \citenamefont
  {Hoffman}(2017)}]{doi:10.1146/annurev-conmatphys-031016-025242}%
  \BibitemOpen
  \bibfield  {author} {\bibinfo {author} {\bibfnamefont {D.}~\bibnamefont
  {Huang}}\ and\ \bibinfo {author} {\bibfnamefont {J.~E.}\ \bibnamefont
  {Hoffman}},\ }\href {\doibase 10.1146/annurev-conmatphys-031016-025242}
  {\bibfield  {journal} {\bibinfo  {journal} {Annual Review of Condensed Matter
  Physics}\ }\textbf {\bibinfo {volume} {8}},\ \bibinfo {pages} {311} (\bibinfo
  {year} {2017})},\ \Eprint
  {http://arxiv.org/abs/https://doi.org/10.1146/annurev-conmatphys-031016-025242}
  {https://doi.org/10.1146/annurev-conmatphys-031016-025242} \BibitemShut
  {NoStop}%
\bibitem [{\citenamefont {Dagotto}(2013)}]{RevModPhys.85.849}%
  \BibitemOpen
  \bibfield  {author} {\bibinfo {author} {\bibfnamefont {E.}~\bibnamefont
  {Dagotto}},\ }\href {\doibase 10.1103/RevModPhys.85.849} {\bibfield
  {journal} {\bibinfo  {journal} {Rev. Mod. Phys.}\ }\textbf {\bibinfo {volume}
  {85}},\ \bibinfo {pages} {849} (\bibinfo {year} {2013})}\BibitemShut
  {NoStop}%
\bibitem [{\citenamefont {Wang}\ \emph {et~al.}(2012)\citenamefont {Wang},
  \citenamefont {Li}, \citenamefont {Zhang}, \citenamefont {Zhang},
  \citenamefont {Zhang}, \citenamefont {Li}, \citenamefont {Ding},
  \citenamefont {Ou}, \citenamefont {Deng}, \citenamefont {Chang},
  \citenamefont {Wen}, \citenamefont {Song}, \citenamefont {He}, \citenamefont
  {Jia}, \citenamefont {Ji}, \citenamefont {Wang}, \citenamefont {Wang},
  \citenamefont {Chen}, \citenamefont {Ma},\ and\ \citenamefont
  {Xue}}]{Wang_2012}%
  \BibitemOpen
  \bibfield  {author} {\bibinfo {author} {\bibfnamefont {Q.-Y.}\ \bibnamefont
  {Wang}}, \bibinfo {author} {\bibfnamefont {Z.}~\bibnamefont {Li}}, \bibinfo
  {author} {\bibfnamefont {W.-H.}\ \bibnamefont {Zhang}}, \bibinfo {author}
  {\bibfnamefont {Z.-C.}\ \bibnamefont {Zhang}}, \bibinfo {author}
  {\bibfnamefont {J.-S.}\ \bibnamefont {Zhang}}, \bibinfo {author}
  {\bibfnamefont {W.}~\bibnamefont {Li}}, \bibinfo {author} {\bibfnamefont
  {H.}~\bibnamefont {Ding}}, \bibinfo {author} {\bibfnamefont {Y.-B.}\
  \bibnamefont {Ou}}, \bibinfo {author} {\bibfnamefont {P.}~\bibnamefont
  {Deng}}, \bibinfo {author} {\bibfnamefont {K.}~\bibnamefont {Chang}},
  \bibinfo {author} {\bibfnamefont {J.}~\bibnamefont {Wen}}, \bibinfo {author}
  {\bibfnamefont {C.-L.}\ \bibnamefont {Song}}, \bibinfo {author}
  {\bibfnamefont {K.}~\bibnamefont {He}}, \bibinfo {author} {\bibfnamefont
  {J.-F.}\ \bibnamefont {Jia}}, \bibinfo {author} {\bibfnamefont {S.-H.}\
  \bibnamefont {Ji}}, \bibinfo {author} {\bibfnamefont {Y.-Y.}\ \bibnamefont
  {Wang}}, \bibinfo {author} {\bibfnamefont {L.-L.}\ \bibnamefont {Wang}},
  \bibinfo {author} {\bibfnamefont {X.}~\bibnamefont {Chen}}, \bibinfo {author}
  {\bibfnamefont {X.-C.}\ \bibnamefont {Ma}}, \ and\ \bibinfo {author}
  {\bibfnamefont {Q.-K.}\ \bibnamefont {Xue}},\ }\href {\doibase
  10.1088/0256-307x/29/3/037402} {\bibfield  {journal} {\bibinfo  {journal}
  {Chinese Physics Letters}\ }\textbf {\bibinfo {volume} {29}},\ \bibinfo
  {pages} {037402} (\bibinfo {year} {2012})}\BibitemShut {NoStop}%
\bibitem [{\citenamefont {Fang}\ \emph {et~al.}(2011)\citenamefont {Fang},
  \citenamefont {Wu}, \citenamefont {Thomale}, \citenamefont {Bernevig},\ and\
  \citenamefont {Hu}}]{PhysRevX.1.011009}%
  \BibitemOpen
  \bibfield  {author} {\bibinfo {author} {\bibfnamefont {C.}~\bibnamefont
  {Fang}}, \bibinfo {author} {\bibfnamefont {Y.-L.}\ \bibnamefont {Wu}},
  \bibinfo {author} {\bibfnamefont {R.}~\bibnamefont {Thomale}}, \bibinfo
  {author} {\bibfnamefont {B.~A.}\ \bibnamefont {Bernevig}}, \ and\ \bibinfo
  {author} {\bibfnamefont {J.}~\bibnamefont {Hu}},\ }\href {\doibase
  10.1103/PhysRevX.1.011009} {\bibfield  {journal} {\bibinfo  {journal} {Phys.
  Rev. X}\ }\textbf {\bibinfo {volume} {1}},\ \bibinfo {pages} {011009}
  (\bibinfo {year} {2011})}\BibitemShut {NoStop}%
\bibitem [{\citenamefont {Lu}\ \emph {et~al.}(2012)\citenamefont {Lu},
  \citenamefont {Fang}, \citenamefont {Tsai}, \citenamefont {Jiang},\ and\
  \citenamefont {Hu}}]{PhysRevB.85.054505}%
  \BibitemOpen
  \bibfield  {author} {\bibinfo {author} {\bibfnamefont {X.}~\bibnamefont
  {Lu}}, \bibinfo {author} {\bibfnamefont {C.}~\bibnamefont {Fang}}, \bibinfo
  {author} {\bibfnamefont {W.-F.}\ \bibnamefont {Tsai}}, \bibinfo {author}
  {\bibfnamefont {Y.}~\bibnamefont {Jiang}}, \ and\ \bibinfo {author}
  {\bibfnamefont {J.}~\bibnamefont {Hu}},\ }\href {\doibase
  10.1103/PhysRevB.85.054505} {\bibfield  {journal} {\bibinfo  {journal} {Phys.
  Rev. B}\ }\textbf {\bibinfo {volume} {85}},\ \bibinfo {pages} {054505}
  (\bibinfo {year} {2012})}\BibitemShut {NoStop}%
\bibitem [{\citenamefont {Yin}\ \emph {et~al.}(2014)\citenamefont {Yin},
  \citenamefont {Haule},\ and\ \citenamefont {Kotliar}}]{yin2014spin}%
  \BibitemOpen
  \bibfield  {author} {\bibinfo {author} {\bibfnamefont {Z.~P.}\ \bibnamefont
  {Yin}}, \bibinfo {author} {\bibfnamefont {K.}~\bibnamefont {Haule}}, \ and\
  \bibinfo {author} {\bibfnamefont {G.}~\bibnamefont {Kotliar}},\ }\href@noop
  {} {\bibfield  {journal} {\bibinfo  {journal} {Nature Physics}\ }\textbf
  {\bibinfo {volume} {10}},\ \bibinfo {pages} {845} (\bibinfo {year}
  {2014})}\BibitemShut {NoStop}%
\bibitem [{\citenamefont {Hu}\ \emph {et~al.}(2013)\citenamefont {Hu},
  \citenamefont {Hao},\ and\ \citenamefont {Wu}}]{hu2013mechanism}%
  \BibitemOpen
  \bibfield  {author} {\bibinfo {author} {\bibfnamefont {J.}~\bibnamefont
  {Hu}}, \bibinfo {author} {\bibfnamefont {N.}~\bibnamefont {Hao}}, \ and\
  \bibinfo {author} {\bibfnamefont {X.}~\bibnamefont {Wu}},\ }\href@noop {}
  {\bibfield  {journal} {\bibinfo  {journal} {arXiv preprint arXiv:1303.2624}\
  } (\bibinfo {year} {2013})}\BibitemShut {NoStop}%
\bibitem [{\citenamefont {Hao}\ and\ \citenamefont
  {Hu}(2014)}]{PhysRevB.89.045144}%
  \BibitemOpen
  \bibfield  {author} {\bibinfo {author} {\bibfnamefont {N.}~\bibnamefont
  {Hao}}\ and\ \bibinfo {author} {\bibfnamefont {J.}~\bibnamefont {Hu}},\
  }\href {\doibase 10.1103/PhysRevB.89.045144} {\bibfield  {journal} {\bibinfo
  {journal} {Phys. Rev. B}\ }\textbf {\bibinfo {volume} {89}},\ \bibinfo
  {pages} {045144} (\bibinfo {year} {2014})}\BibitemShut {NoStop}%
\bibitem [{\citenamefont {Du}\ \emph {et~al.}(2018)\citenamefont {Du},
  \citenamefont {Yang}, \citenamefont {Altenfeld}, \citenamefont {Gu},
  \citenamefont {Yang}, \citenamefont {Eremin}, \citenamefont {Hirschfeld},
  \citenamefont {Mazin}, \citenamefont {Lin}, \citenamefont {Zhu} \emph
  {et~al.}}]{du2018sign}%
  \BibitemOpen
  \bibfield  {author} {\bibinfo {author} {\bibfnamefont {Z.}~\bibnamefont
  {Du}}, \bibinfo {author} {\bibfnamefont {X.}~\bibnamefont {Yang}}, \bibinfo
  {author} {\bibfnamefont {D.}~\bibnamefont {Altenfeld}}, \bibinfo {author}
  {\bibfnamefont {Q.}~\bibnamefont {Gu}}, \bibinfo {author} {\bibfnamefont
  {H.}~\bibnamefont {Yang}}, \bibinfo {author} {\bibfnamefont {I.}~\bibnamefont
  {Eremin}}, \bibinfo {author} {\bibfnamefont {P.~J.}\ \bibnamefont
  {Hirschfeld}}, \bibinfo {author} {\bibfnamefont {I.~I.}\ \bibnamefont
  {Mazin}}, \bibinfo {author} {\bibfnamefont {H.}~\bibnamefont {Lin}}, \bibinfo
  {author} {\bibfnamefont {X.}~\bibnamefont {Zhu}},  \emph {et~al.},\
  }\href@noop {} {\bibfield  {journal} {\bibinfo  {journal} {Nature Physics}\
  }\textbf {\bibinfo {volume} {14}},\ \bibinfo {pages} {134} (\bibinfo {year}
  {2018})}\BibitemShut {NoStop}%
\bibitem [{\citenamefont {Fan}\ \emph {et~al.}(2015)\citenamefont {Fan},
  \citenamefont {Zhang}, \citenamefont {Liu}, \citenamefont {Yan},
  \citenamefont {Ren}, \citenamefont {Peng}, \citenamefont {Xu}, \citenamefont
  {Xie}, \citenamefont {Hu}, \citenamefont {Zhang} \emph
  {et~al.}}]{fan2015plain}%
  \BibitemOpen
  \bibfield  {author} {\bibinfo {author} {\bibfnamefont {Q.}~\bibnamefont
  {Fan}}, \bibinfo {author} {\bibfnamefont {W.}~\bibnamefont {Zhang}}, \bibinfo
  {author} {\bibfnamefont {X.}~\bibnamefont {Liu}}, \bibinfo {author}
  {\bibfnamefont {Y.}~\bibnamefont {Yan}}, \bibinfo {author} {\bibfnamefont
  {M.}~\bibnamefont {Ren}}, \bibinfo {author} {\bibfnamefont {R.}~\bibnamefont
  {Peng}}, \bibinfo {author} {\bibfnamefont {H.}~\bibnamefont {Xu}}, \bibinfo
  {author} {\bibfnamefont {B.}~\bibnamefont {Xie}}, \bibinfo {author}
  {\bibfnamefont {J.}~\bibnamefont {Hu}}, \bibinfo {author} {\bibfnamefont
  {T.}~\bibnamefont {Zhang}},  \emph {et~al.},\ }\href@noop {} {\bibfield
  {journal} {\bibinfo  {journal} {Nature Physics}\ }\textbf {\bibinfo {volume}
  {11}},\ \bibinfo {pages} {946} (\bibinfo {year} {2015})}\BibitemShut
  {NoStop}%
\bibitem [{\citenamefont {Wu}\ \emph {et~al.}(2016)\citenamefont {Wu},
  \citenamefont {Liang}, \citenamefont {Fan},\ and\ \citenamefont
  {Hu}}]{wu2016nematic}%
  \BibitemOpen
  \bibfield  {author} {\bibinfo {author} {\bibfnamefont {X.}~\bibnamefont
  {Wu}}, \bibinfo {author} {\bibfnamefont {Y.}~\bibnamefont {Liang}}, \bibinfo
  {author} {\bibfnamefont {H.}~\bibnamefont {Fan}}, \ and\ \bibinfo {author}
  {\bibfnamefont {J.}~\bibnamefont {Hu}},\ }\href@noop {} {\bibfield  {journal}
  {\bibinfo  {journal} {arXiv preprint arXiv:1603.02055}\ } (\bibinfo {year}
  {2016})}\BibitemShut {NoStop}%
\bibitem [{\citenamefont {Ahn}\ and\ \citenamefont
  {Yang}(2021)}]{PhysRevB.103.184502}%
  \BibitemOpen
  \bibfield  {author} {\bibinfo {author} {\bibfnamefont {J.}~\bibnamefont
  {Ahn}}\ and\ \bibinfo {author} {\bibfnamefont {B.-J.}\ \bibnamefont {Yang}},\
  }\href {\doibase 10.1103/PhysRevB.103.184502} {\bibfield  {journal} {\bibinfo
   {journal} {Phys. Rev. B}\ }\textbf {\bibinfo {volume} {103}},\ \bibinfo
  {pages} {184502} (\bibinfo {year} {2021})}\BibitemShut {NoStop}%
\end{thebibliography}%

%
%
%

\appendix

\begin{widetext}

\subsection{Mirror symmetry in space group $P4/nmm$}
In the main text, we show that the band degeneracies along the $\Sigma_{ \text{Y} }$ and $\Sigma_{ \text{G} }$ lines are very different, based on the analysis of the group structure of $P4/nmm$. Here, we provide a detail calculation by decoupling the normal-state lattice Hamiltonian in the main text according to the mirror symmetry $\{ M_y | {\bf 0} \}$.

We first rotate the spin to  the $y$ direction. Namely, we do a unitary transformation $\mathcal{H}_0^\prime = e^{ \frac{-i\pi}{4} s_1 } \mathcal{H}_0 e^{ \frac{i\pi}{4} s_1 }$. In $\mathcal{H}_0^\prime$, the spin $\uparrow\downarrow$ are along the $y$ direction and $p_{x/y}$ are eigenstates of $\{ M_y | {\bf 0} \}$. Therefore, in general $( c_{Ax\uparrow}, c_{Ay\downarrow}, c_{Bx\uparrow}, c_{By\downarrow} )$ ($c_{Ax\uparrow}$ is the annihilation operator for the spin-$\uparrow$ $p_x$-orbital electron in the A sublattice) would be in a mirror invariant subspace, while their time reversal partners  are in the other mirror subspace.  On the line $\Sigma_{\text{G}}$ for group $P4/nmm$,  we can write down $\mathcal{H}_0^\prime$  as
\begin{eqnarray}\label{H_G}
\small { \left(
	\begin{array}{cccccccc}
		2t\cos k_x - 2\lambda_R \sin k_x  & 0 & 0 & \frac{\lambda}{2} & 4t_1\cos\frac{k_x}{2} & 0 & 0 & 0   \\
		0 & 2t & -\frac{\lambda}{2} & 0 & 0 & 4t_1\cos\frac{k_x}{2} & 0 & 0   \\
        0 & -\frac{\lambda}{2} & 2t\cos k_x + 2\lambda_R \sin k_x & 0 & 0 & 0 & 4t_1\cos\frac{k_x}{2} & 0   \\
        \frac{\lambda}{2}  & 0 & 0 & 2t & 0 & 0 & 0 & 4t_1\cos\frac{k_x}{2}   \\
        4t_1\cos\frac{k_x}{2} & 0 & 0 & 0 & 2t\cos k_x + 2\lambda_R \sin k_x & 0 & 0 & \frac{\lambda}{2}   \\
        0 & 4t_1\cos\frac{k_x}{2} & 0 & 0 & 0 & 2t & -\frac{\lambda}{2} & 0   \\
        0 & 0 & 4t_1\cos\frac{k_x}{2} & 0 & 0 & -\frac{\lambda}{2} & 2t\cos k_x - 2\lambda_R \sin k_x & 0   \\
        0 & 0 & 0 & 4t_1\cos\frac{k_x}{2} & \frac{\lambda}{2} & 0 & 0 & 2t   \\
	\end{array}
	\right) },
\end{eqnarray}
with the basis is $( c_{Ax\uparrow}, c_{Ay\uparrow}, c_{Ax\downarrow}, c_{Ay\downarrow}, c_{Bx\uparrow}, c_{By\uparrow}, c_{Bx\downarrow}, c_{By\downarrow} )$. The situation is very different on line $\Sigma_{\text{Y}}$. Under the same basis with the Hamiltonian in Eq.\eqref{H_G}, $\mathcal{H}_0^\prime$ on $\Sigma_{\text{Y}}$ takes the form
\begin{eqnarray}\label{H_Y}
\small { \left(
	\begin{array}{cccccccc}
		2t\cos k_x - 2\lambda_R \sin k_x  & 0 & 0 & \frac{\lambda}{2} & 0 & -4t_2\sin\frac{k_x}{2} & 0 & 0   \\
		0 & 2t & -\frac{\lambda}{2} & 0 & -4t_2\sin\frac{k_x}{2} & 0 & 0 & 0   \\
        0 & -\frac{\lambda}{2} & 2t\cos k_x + 2\lambda_R \sin k_x & 0 & 0 & 0 & 0 & -4t_2\sin\frac{k_x}{2}   \\
        \frac{\lambda}{2}  & 0 & 0 & 2t & 0 & 0 & -4t_2\sin\frac{k_x}{2} & 0   \\
        0 & -4t_2\sin\frac{k_x}{2} & 0 & 0 & 2t\cos k_x + 2\lambda_R \sin k_x & 0 & 0 & \frac{\lambda}{2}   \\
        -4t_2\sin\frac{k_x}{2} & 0 & 0 & 0 & 0 & 2t & -\frac{\lambda}{2} & 0   \\
        0 & 0 & 0 & -4t_2\sin\frac{k_x}{2} & 0 & -\frac{\lambda}{2} & 2t\cos k_x - 2\lambda_R \sin k_x & 0   \\
        0 & 0 & -4t_2\sin\frac{k_x}{2} & 0 & \frac{\lambda}{2} & 0 & 0 & 2t   \\
	\end{array}
	\right) },
\end{eqnarray}
In this case, $( c_{Ax\uparrow}, c_{Ay\downarrow}, c_{By\uparrow}, c_{Bx\downarrow} )$ are in a mirror invariant subspace.

The above difference can be straightforwardly understood by considering the Fourier transform of the real-space basis
\begin{eqnarray}\label{Fourier}
| \phi_A ({\bf k}) \rangle = \sum_j e^{i {\bf k} \cdot {\bf R}_A^j } | \phi_A ({\bf R}_A^j) \rangle, \quad | \phi_B ({\bf k}) \rangle = \sum_j e^{i {\bf k} \cdot {\bf R}_B^j } | \phi_B ({\bf R}_B^j) \rangle,
\end{eqnarray}
where ${\bf R}_{A/B}^j$ is the position of the A/B site in the $j$-th unit cell, with ${\bf R}_A^j - {\bf R}_B^j = {\bf \tau_0}$. Applying the mirror symmetry $(k_y \rightarrow -k_y)$, we have
\begin{eqnarray}\label{mirror_transform}
k_y &=& \pi: \  \{ M_y | {\bf 0} \} | \phi_A ({\bf k}) \rangle = \sum_j e^{i {\bf k} \cdot {\bf R}_A^j } m_A | \phi_A ({\bf R}_A^j) \rangle, \quad \{ M_y | {\bf 0} \} | \phi_B ({\bf k}) \rangle = -\sum_j e^{i {\bf k} \cdot {\bf R}_B^j } m_B | \phi_B ({\bf R}_B^j) \rangle, \nonumber \\
k_y &=& 0: \  \{ M_y | {\bf 0} \} | \phi_A ({\bf k}) \rangle = \sum_j e^{i {\bf k} \cdot {\bf R}_A^j } m_A | \phi_A ({\bf R}_A^j) \rangle, \quad \{ M_y | {\bf 0} \} | \phi_B ({\bf k}) \rangle = \sum_j e^{i {\bf k} \cdot {\bf R}_B^j } m_B | \phi_B ({\bf R}_B^j) \rangle,
\end{eqnarray}
with $m_{A/B}$ the mirror eigenvalues. In the equation, we have considered the fact that $R_{A/B, y}^j \cdot k_y = j \pi$ at the Brillouin zone boundary (definition of $k_y$). The results in Eq.\eqref{mirror_transform} explain the difference between the mirror invariant subspaces in Eqs.\eqref{H_G}\eqref{H_Y}.

\subsection{Symmetry constraints on the winding number}
In this part, we show how the symmetries constrain the winding number. As mentioned in the main text, for a 1D system $\mathcal{H}$ with chiral symmetry $\mathcal{C}$, its topological property can be characterized by the winding number $w$. If the system has additional symmetry $g$ satisfying $g \mathcal{H}(k) g^{-1} = \mathcal{H}({\bf g} k)$, the symmetry has constraint on $w$ as follows\cite{PhysRevB.103.184502}
\begin{eqnarray}\label{winding_constraint}
w &=& \int_{-\pi}^\pi \frac{dk}{2\pi} tr[ \mathcal{C} \mathcal{H}^{-1}(k) \partial_{k} \mathcal{H}(k) ] \nonumber \\
&=& \int_{-\pi}^{\pi} \frac{dk}{2\pi} tr[ \mathcal{C} \left( g \mathcal{H}({\bf g}^{-1}k) g^{-1} \right)^{-1} \partial_{k} \left( g \mathcal{H}({\bf g}^{-1}k) g^{-1} \right) ] \nonumber \\
&=& \int_{-\pi}^\pi \frac{dk}{2\pi} tr[ g^{-1} \mathcal{C} g \mathcal{H}^{-1}({\bf g}^{-1}k) \partial_{k} \mathcal{H}({\bf g}^{-1}k) ] \nonumber \\
&=& \int_{-g^{-1}\pi}^{g\pi} \frac{d({\bf g}k)}{2\pi} tr[ g^{-1} \mathcal{C} g \mathcal{H}^{-1}(k) \partial_{{\bf g}k} \mathcal{H}(k) ] \nonumber \\
&=& \int_{-\pi}^{\pi} \frac{dk}{2\pi} tr[ g^{-1} \mathcal{C} g \mathcal{H}^{-1}(k) \partial_{k} \mathcal{H}(k) ] \cdot det({\bf g}) \nonumber \\
&=& \int_{-\pi}^{\pi} \frac{dk}{2\pi} tr[ \mathcal{C} \mathcal{H}^{-1}(k) \partial_{k} \mathcal{H}(k) ] \cdot det({\bf g}),
\end{eqnarray}
where we have  used the property, $[\mathcal{C}, g] = 0$, for  the s-wave  superconductivity.

Now we consider the winding number in each of the $\{ M_y | {\bf 0} \}$ subspaces on the $\Sigma_{ \text{Y}/\text{G} }$ line, where the bands are always two-fold degenerate. Based on the symmetry constraints, specifically
\begin{eqnarray}\label{SM_commutation1}
(\{ I | {\bf \tau_0} \} \Theta) \{ M_z | {\bf \tau_0} \} | \varphi({\bf k}) \rangle &=& e^{ -2i {\bf k} \cdot {\bf \tau_0} }\{ M_z | {\bf \tau_0} \} (\{ I | {\bf \tau_0} \} \Theta) | \varphi({\bf k}) \rangle, \nonumber \\
(\{ I | {\bf \tau_0} \} \Theta) \{ M_y | {\bf 0} \} | \varphi({\bf k}) \rangle &=& e^{ik_y} \{ M_y | {\bf 0} \} (\{ I | {\bf \tau_0} \} \Theta) | \varphi({\bf k}) \rangle, \nonumber \\
\{ M_z | {\bf \tau_0} \} \{ M_y | {\bf 0} \} | \varphi({\bf k}) \rangle &=& -e^{ik_y} \{ M_y | {\bf 0} \} \{ M_z | {\bf \tau_0} \} | \varphi({\bf k}) \rangle,
\end{eqnarray}
namely
\begin{eqnarray}\label{SM_commutation2}
k_y = \pi: \ & [ \{ I | {\bf \tau_0} \} \Theta, \{ M_y | {\bf 0} \} ]_+ = 0, \
[ \{ M_z | {\bf \tau_0} \}, \{ M_y | {\bf 0} \} ]_- = 0, \nonumber \\
k_y = 0: \ & [ \{ I | {\bf \tau_0} \} \Theta, \{ M_y | {\bf 0} \} ]_- = 0, \
[ \{ M_z | {\bf \tau_0} \}, \{ M_y | {\bf 0} \} ]_+ = 0,
\end{eqnarray}
we can obtain the following conclusions.

\noindent
(i) On $\Sigma_{\text{Y}}$ the two degenerate states are in the same mirror subspace which can be further decoupled by $\{ M_z | {\bf \tau_0} \}$. Since the two states are related by the symmetry $\{ I | {\bf \tau_0} \} \Theta$, we have $w_{\text{Y}}^{m_y, m_z} = w_{\text{Y}}^{m_y, -m_z}$, with $m_y = \pm i$ and $m_z = \pm i e^{ i {\bf k} \cdot {\bf \tau_0} }$ are the eigenvalues of $\{ M_y | {\bf 0} \}$ and $\{ M_z | {\bf \tau_0} \}$ respectively. Considering the constraint of the time reversal symmetry,  we have $w_{\text{Y}}^{+i} = -w_{\text{Y}}^{-i} = 2 \mathcal{Z}$;

\noindent
(ii) On $\Sigma_{\text{G}}$ we take the inversion symmetry $\{ I | {\bf \tau_0} \}$ into consideration, which satisfies $\{ M_y | {\bf 0} \} \{ I | {\bf \tau_0} \} | \varphi({\bf k}) \rangle = e^{ i k_y } \{ I | {\bf \tau_0} \} \{ M_y | {\bf 0} \} | \varphi({\bf k}) \rangle = \{ I | {\bf \tau_0} \} \{ M_y | {\bf 0} \} | \varphi({\bf k}) \rangle$. Obviously, $\{ I | {\bf \tau_0} \}$ maps a state at ${\bf k}$ to a state at $-{\bf k}$ in the same mirror subspace. According to Eq.\eqref{winding_constraint},  we have $w_{\text{G}}^{+i} = -w_{\text{G}}^{-i} = 0$.

\subsection{Topological trivialness in centrosymmetric symmorphic superconductors with the s-wave ($A_{1g}$) pairing}
In this part, we show that in centrosymmetric superconductors respecting symmorphic groups, if the pairing order is in the $A_{1g}$ channel, the superconductivity is always topologically trivial with respect to (i) the topological indices in the AZ classification, and (ii) the mirror protected topological indices (the mirror Chern number and mirror winding number).

\subsubsection{Without crystalline symmetry}
We begin with the topological indices in the AZ classification. In the 1D and 3D cases, the class-D\uppercase\expandafter{\romannumeral3} superconductors are featured by the $\mathcal{Z}$ indices, which are the 1D and 3D winding numbers defined according to the chiral symmetry $\mathcal{C}$. We label the inversion symmetry with $\mathcal{I}$. As mentioned, for the $A_{1g}$ superconductivity $[\mathcal{C}, \mathcal{I}] = 0$. Moreover, since ${\bf \mathcal{I} } {\bf k} = -{\bf k}$, namely $det( {\bf \mathcal{I} } ) = -1$. Hence, in 1D case,  $w_{1D} = 0$ according to Eq.\eqref{winding_constraint}. The 3D case can be analyzed similarly. In the presence of the inversion symmetry, the 3D winding number is confined as
\begin{eqnarray}\label{3D_winding}
w_{3D} &=& \int \frac{d^3k}{48\pi^2} \epsilon^{\alpha\beta\gamma} tr[ \mathcal{C} \mathcal{H}^{-1}(k) \partial_{k_\alpha} \mathcal{H}(k) \mathcal{H}^{-1}(k) \partial_{k_\beta} \mathcal{H}(k) \mathcal{H}^{-1}(k) \partial_{k_\gamma} \mathcal{H}(k) ] \nonumber \\
&=& \int \frac{d^3k}{48\pi^2} \epsilon^{\alpha\beta\gamma} tr[ \mathcal{C} \left( \mathcal{I} \mathcal{H}(-{\bf k}) \mathcal{I}^{-1} \right)^{-1} \partial_{k_\alpha} \left( \mathcal{I} \mathcal{H}(-{\bf k}) \mathcal{I}^{-1} \right)  \left( \mathcal{I} \mathcal{H}(-{\bf k}) \mathcal{I}^{-1} \right)^{-1} \nonumber \\
&& \qquad \qquad \quad  \partial_{k_\beta} \left( \mathcal{I} \mathcal{H}(-{\bf k}) \mathcal{I}^{-1} \right) \left( \mathcal{I} \mathcal{H}(-{\bf k}) \mathcal{I}^{-1} \right)^{-1} \partial_{k_\gamma} \left( \mathcal{I} \mathcal{H}(-{\bf k}) \mathcal{I}^{-1} \right) ], \nonumber \\
&=& \int \frac{d^3k}{48\pi^2} \epsilon^{\alpha\beta\gamma} tr[ \mathcal{C} \mathcal{H}^{-1}(k) \partial_{k_\alpha} \mathcal{H}(k) \mathcal{H}^{-1}(k) \partial_{k_\beta} \mathcal{H}(k) \mathcal{H}^{-1}(k) \partial_{k_\gamma} \mathcal{H}(k) ] \cdot det({\bf \mathcal{I} }), \nonumber\\
&=& -w_{3D} = 0,
\end{eqnarray}
where $[\mathcal{C}, \mathcal{I}] = 0$ and ${\bf \mathcal{I} } {\bf k} = -{\bf k}$ have been taken into consideration. Therefore, in 1D and 3D,  the inversion symmetry demands the $A_{1g}$ superconductivity to be topologically trivial.

For the 2D case, the superconductor is featured by a $\mathcal{Z}_2$ index similar to that in a time reversal symmetry protected topological insulator. In general, it is hard to calculate the $\mathcal{Z}_2$ index directly. However,  due to the inversion symmetry, we can calculate the $\mathcal{Z}_2$ index based on the parity criterion
\begin{eqnarray}\label{Z2}
\mathcal{Z}_2: \quad (-1)^\nu = \prod_i \prod_{m=1}^N \xi_{2m}(\Gamma_i),
\end{eqnarray}
where $\Gamma_i$ ($i = 1, 2, 3, 4$) are the four time reversal invariant points in the 2D Brillouin zone, and $\xi_{2m}$ is the parity of the $2m$-th negative-energy states (the time reversal symmetry requires that all of the states appear in the form of Kramers' pairs, and each Kramers' pair shares the same parity). We take the chiral symmetry $\mathcal{C}$ into consideration, which is unitary and satisfies $\mathcal{C} \mathcal{H}({\bf k}) \mathcal{C}^{-1} = -\mathcal{H}({\bf k})$. We can realize that $\mathcal{C}$ maps a state ($E({\bf k})$, $\xi_0$) (with $E({\bf k})$ the eigen energy and $\xi_0$ the parity) to a state (-$E({\bf k})$, $\xi_0$), considering$[ \mathcal{C}, \mathcal{I} ] = 0$. Based on this relation,  for the $A_{1g}$ superconductivity, the $\mathcal{Z}_2$ index in Eq.\eqref{Z2} can only take one value. Namely, it is always topologically trivial.

\subsubsection{In the presence of mirror symmetry}
We turn to the situation where there is mirror symmetry in a centrosymmetric class-D\uppercase\expandafter{\romannumeral3} superconductor, and consider the mirror-protected topological indices, mainly the mirror Chern number and mirror winding number.

We first consider the condition where there are mirror invariant lines in the Brillouin zone. Specifically, two cases are included: (i) In 1D, the system is parallel to the mirror plane; (ii) In 2D, the mirror plane is normal to the system (for simplicity we consider a square lattice with mirror symmetry $\mathcal{M}_y$). In these two cases, we study the mirror-protected winding numbers, and the analysis is similar to the case of $\Sigma_{\text{G}}$ line in the main text. The winding number in each of the mirror invariant subspaces is always trivial, namely $w_{1D}^{+i} = w_{1D}^{-i} = 0$ and $w_{2D}^{+i}( k_y = 0 ) = w_{2D}^{-i}( k_y = 0 ) = w_{2D}^{+i}( k_y = \pi ) = w_{2D}^{-i}( k_y = \pi ) = 0$.

Then, we consider the condition where there is mirror invariant planes in 3D superconductors (for simplicity we consider the mirror symmetry $\mathcal{M}_z$ in a tetragonal lattice). Within the mirror invariant planes, the mirror Chern numbers are well-defined. As pointed out in Ref.\cite{PhysRevLett.111.056403}, the Chern numbers are always zeros in each of the mirror subspaces, $C_{+i} = C_{-i} = 0$, because of the chiral symmetry. Moreover, within the mirror invariant planes the mirror-protected winding number can be considered along the time reversal invariant lines in the $k_z = 0/\pi$. For instance, the mirror-protected winding number is well-defined on lines $(k_x, k_y^0, k_z^0)$ and $(k_x^0, k_y, k_z^0)$, with $k_{x/y}^0 = 0, \pi$ and $-\pi \leq k_{x/y} \leq \pi$. Because of the inversion symmetry, these mirror-protected winding numbers are always zero, which is similar to the case of $\Sigma_{\text{G}}$ line in the main text. The above analysis is also true for   2D superconductors where  the mirror plane is parallel to the system.

\subsection{Edge theory for the corner modes}
In this part, we derive an effective edge theory to explain that the appearance of corner Majorana modes is guaranteed by the mirror symmetry, namely the topological superconductivity in the main text is an intrinsic second-order one.

\begin{figure}[!htbp]
	\centering
	\includegraphics[width=0.4\linewidth]{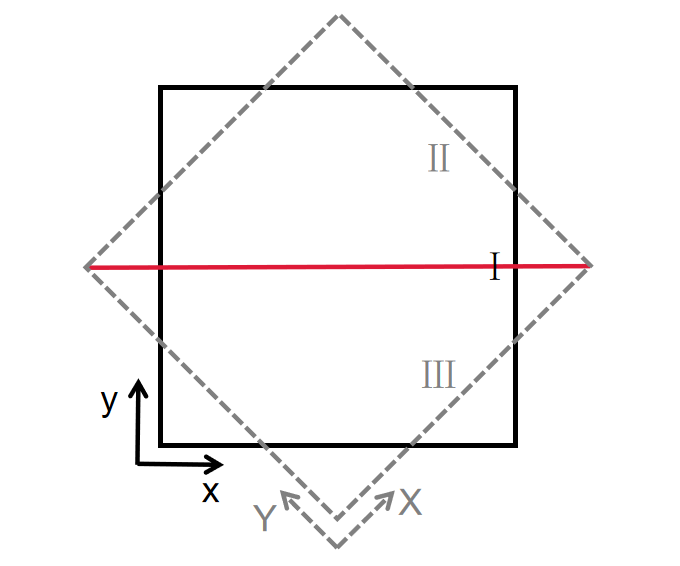}
	\caption{\label{SM_edge} (color online) Sketch of two different open-boundary conditions. The square in black solid lines shows the open boundaries in the $(10)$ and $(01)$ directions, while the square in gray dashed lines shows the open boundaries in the $(11)$ and $(1 \overline{1} )$ directions. The mirror symmetry is implied by the red line in the figure.
	}
\end{figure}

On the $(10)$ edge, the mirror symmetry $\mathcal{M}_y = \{ M_y | {\bf 0} \}$ is preserved. In addition to $\mathcal{M}_y$, the particle-hole symmetry $\mathcal{P}$ and the time reversal symmetry $\Theta$ are also preserved. The chiral symmetry $\mathcal{C} = \mathcal{P} \Theta$ also exists on the edge. An effective theory describing the Dirac cones on the edge, can be   derived   based on the constraints of the above four symmetries. These symmetries constrain the edge theory on the $(10)$ edge as
\begin{eqnarray}\label{edge_symmetry1}
\Theta \mathcal{H}_{eff}(k_y) \Theta^{-1} = \mathcal{H}(-k_y), \quad
\mathcal{P} \mathcal{H}_{eff}(k_y) \mathcal{P}^{-1} = -\mathcal{H}(-k_y), \quad
\mathcal{C} \mathcal{H}_{eff}(k_y) \mathcal{C}^{-1} = -\mathcal{H}(k_y), \quad
\mathcal{M} \mathcal{H}_{eff}(k_y) \mathcal{M}^{-1} = \mathcal{H}(-k_y).
\end{eqnarray}
Moreover, the symmetries satisfy
\begin{eqnarray}\label{edge_symmetry2}
[\Theta, \mathcal{M}]_- = [\mathcal{P}, \mathcal{M}]_- = [\mathcal{C}, \mathcal{M}]_- = 0.
\end{eqnarray}
According to these constraints, with a proper gauge choice the matrix form of the above symmetry operators can be chosen as $\Theta = i s_2 \kappa_0 K$, $\mathcal{P} = i s_2 \kappa_2 K$, $\mathcal{C} = s_0 \kappa_2$ and $\mathcal{M} = i s_2 \kappa_0$, with $K$ the complex conjugate operation. Here, $s_i$ are the Pauli matrices labeling the spin degree of freedom and $\kappa_i$ the Pauli matrices labeling the remaining degree for the two Dirac cones on the edge. Correspondingly, the effective theory on edge \uppercase\expandafter{\romannumeral1} in Fig.\ref{SM_edge} can take the following form $\mathcal{H}_{eff} = v k_y ( \alpha s_1 \kappa_1 + \beta s_3 \kappa_3)$ or $\mathcal{H}_{eff} = v k_y ( \alpha s_3 \kappa_1 + \beta s_1 \kappa_3 )$, with $v$ the Fermi velocity. For simplicity, we take $\mathcal{H}_{eff} = v k_y s_3 \kappa_3$ in the following (the other cases can be analyzed similarly).

Then, we bend the edge \uppercase\expandafter{\romannumeral1} and make the bent edge evolve into the edge \uppercase\expandafter{\romannumeral2} and \uppercase\expandafter{\romannumeral3} in Fig.\ref{SM_edge} gradually. In this progress, the gapless modes on the edge \uppercase\expandafter{\romannumeral2} (\uppercase\expandafter{\romannumeral3}) gain a mass and are gapped out, as the mirror symmetry can not be maintained on each of the edges. Considering the symmetry constraints in Eq.\eqref{edge_symmetry1}, we have the mass term as $m_{ \uppercase\expandafter{\romannumeral2} / \uppercase\expandafter{\romannumeral3} } ( \alpha s_3 \kappa_1+ \beta s_1 \kappa_3 )$, with $m_{ \uppercase\expandafter{\romannumeral2} / \uppercase\expandafter{\romannumeral3} }$ the mass term on edge \uppercase\expandafter{\romannumeral2}/\uppercase\expandafter{\romannumeral3}. For simplicity, we take the mass term $m_{ \uppercase\expandafter{\romannumeral2} / \uppercase\expandafter{\romannumeral3} } \beta s_1 \kappa_3$  and  the effective theories on these two edges  can be written as
\begin{eqnarray}\label{edge_theory}
\mathcal{H}_{eff, \uppercase\expandafter{\romannumeral2}} = v k_Y s_3 \kappa_3 + m_{ \uppercase\expandafter{\romannumeral2} } s_1 \kappa_3, \quad
\mathcal{H}_{eff, \uppercase\expandafter{\romannumeral3}} = v k_X s_3 \kappa_3 + m_{ \uppercase\expandafter{\romannumeral3} } s_1 \kappa_3.
\end{eqnarray}
Moreover, since edge \uppercase\expandafter{\romannumeral2} and \uppercase\expandafter{\romannumeral3} are related by the mirror reflection $\mathcal{M}$, the effective theories on the two edges satisfy $\mathcal{M} \mathcal{H}_{eff, \uppercase\expandafter{\romannumeral2}} \mathcal{M}^{-1} = \mathcal{H}_{eff, \uppercase\expandafter{\romannumeral3}}$, As  ${\bf \mathcal{M}} k_X = -k_Y$, we have  $m_{ \uppercase\expandafter{\romannumeral2} } = -m_{ \uppercase\expandafter{\romannumeral3} }$.

The theory in Eq.\eqref{edge_theory} can be written in an instructive form by treating the gapless part as propagating modes along the edges of a finite size system, $\mathcal{H}_{eff} = v k s_3 \kappa_3 + m_{ r } s_1 \kappa_3$ with $m_{ r }$ changing size at the intersection between the $(11)$ and $(1 \overline{1})$ edges (assuming at $r = 0$). In this form, $\mathcal{H}_{eff}$ is just a Dirac theory with a mass domain at $r = 0$, which naturally leads to a pair of Majorana modes localized at $r = 0$, $i.e.$ the corner.

\subsection{Calculations for the iron-selenide superconductors}
In this part, we derive the topological superconductivity for iron-chalcogenides. We consider the following  Hamiltonian that captures the band structure of iron-based superconductors,
\begin{eqnarray}\label{SM_iron1}
\mathcal{H}_0 = \mathcal{H}_{\text{TB}} + \mathcal{H}_{\text{soc}}.
\end{eqnarray}
where $\mathcal{H}_{\text{TB}}$ and $\mathcal{H}_{\text{soc}}$ are the tight-binding part and the spin-orbit coupling part respectively. For the tight-binding part, we adopt the Hamiltonian in Ref.\cite{wu2016nematic} in which all five $d$ orbitals of the Fe atoms are taken into account, and utilize the parameters in Table.\ref{iron_TB} to fit the band structures of a single layer FeSe shown in Fig.\ref{fig_iron_SM}(a). For the spin-orbit coupling part, we merely consider the atomic spin-orbit coupling of the $d$ orbitals, namely $\mathcal{H}_{\text{soc}} = \lambda {\bf L} \cdot {\bf s}$ with ${\bf L}/{\bf s}$ the orbital/spin angular momentum. Fig.\ref{fig_iron_SM}(b) shows the bands in the presence of the spin-orbit coupling ($\lambda = 40$ meV). In the following, we set the chemical potential to be $\mu = 118.3$ meV (the doping level is about $0.09$ electron per Fe), and the corresponding Fermi surfaces are presented in Fig.\ref{fig_iron_SM}(c)(d).

\begin{table}[]\centering
\caption{\label{iron_TB} Hopping parameters for the monolayer FeSe. The onsite energies of the d orbitals are: $\epsilon_1 = 0.1754$, $\epsilon_3 = -0.3576$, $\epsilon_4 = 0.0904$, $\epsilon_5 = -0.2776$. Notice that we adopt the same notations with those in Ref.\cite{wu2016nematic}, and all the parameters are in the unit of eV.   }
\begin{tabular}{p{1.8cm}<{\centering} p{1.8cm}<{\centering} p{1.8cm}<{\centering} p{1.8cm}<{\centering} p{1.8cm}<{\centering} p{1.8cm}<{\centering} p{1.8cm}<{\centering} p{1.8cm}<{\centering} p{1.8cm}<{\centering} }
\hline\hline
$t_i^{mn}$      & i=x     & i=y    & i=xy   & i=xx   & i=yy   & i=xxy  & i=xyy  & i=xxyy \\ \hline
mn=11 & -0.1514 & -0.4059 & 0.225  & 0.002  & -0.036 & -0.019 & 0.014  & 0.024  \\
mn=33 & -0.4584 &         & -0.070 & -0.013 &        &        &        & 0.012  \\
mn=44 & -0.0704 &         & 0.012  & 0.002  &        & 0.019  &        & -0.024 \\
mn=55 &         &         & 0.013  & -0.014 &        & -0.006 &        & -0.011 \\
mn=12 &         &         & 0.103  &        &        & -0.011 &        & 0.032  \\
mn=13 & -0.473  &         & -0.089 &        & 0.011  & 0.018  & -0.006 &        \\
mn=14 & -0.2736 &         & 0.053  & -0.001 &        & -0.006 &        & -0.009 \\
mn=15 & -0.200  &         & -0.130 &        & 0.009  & 0.009  & 0.011  & -0.012 \\
mn=34 &         &         &        &        &        & 0.012  &        &        \\
mn=35 & 0.401   &         &        & -0.023 &        & 0.006  &        &        \\
mn=45 &         &         & -0.113 &        &        &        &        & 0.011  \\ \hline\hline
\end{tabular}
\end{table}

\begin{figure}[!htbp]
	\centering
	\includegraphics[width=0.75\linewidth]{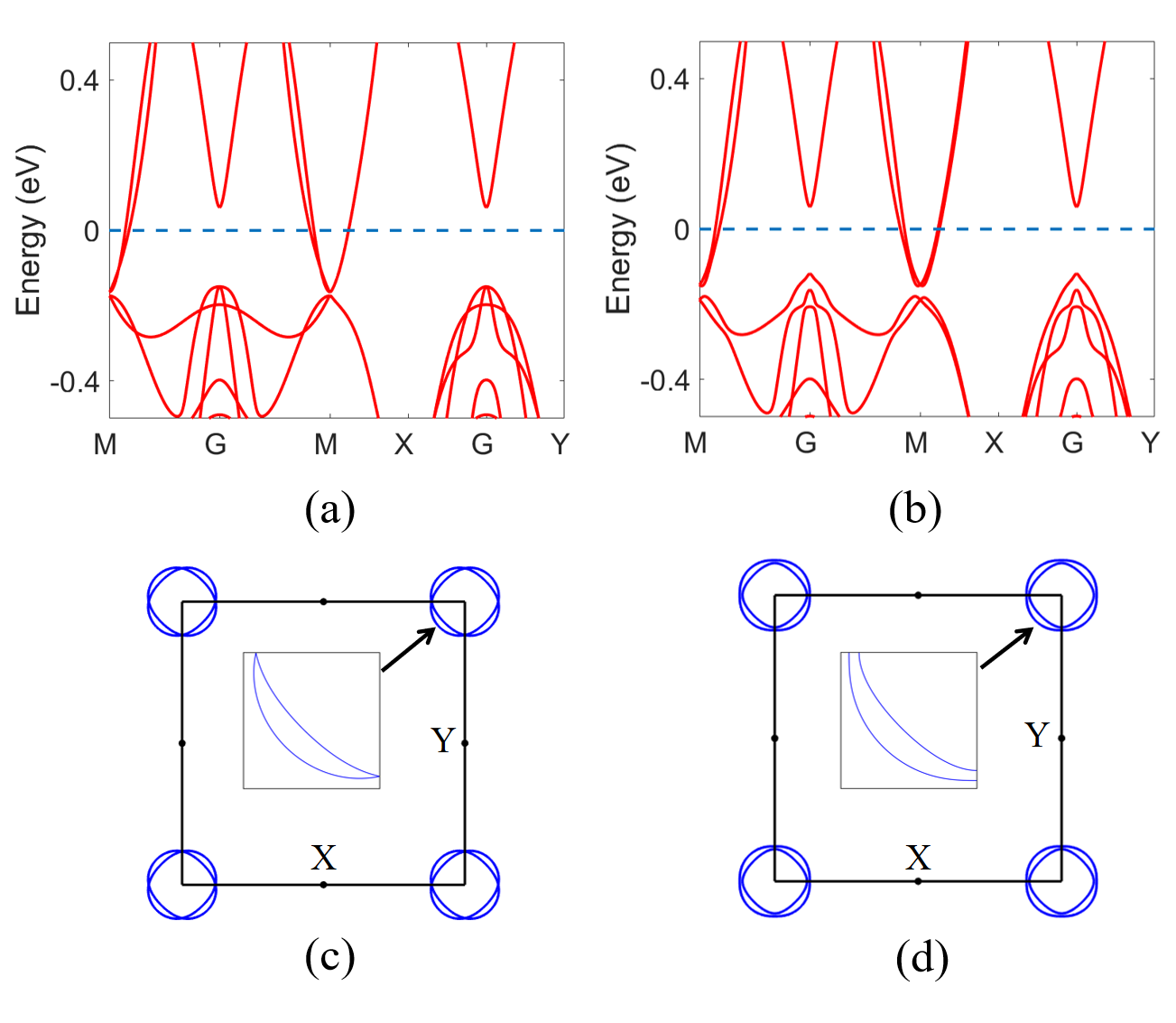}
	\caption{\label{fig_iron_SM} (color online) (a) and (b) show the normal-state band structures of the monolayer FeSe in the absence and presence of the spin-orbit coupling respectively. The corresponding Fermi surfaces are shown in (c) and (d), with the insets showing the details. The chemical potential is set to be $\mu = 118.3$ meV and the spin-orbit coupling $\lambda = 40$ meV here.
	}
\end{figure}

For the superconducting part, we design a general superconducting gap form which includes the onsite and NN intra-sublattice intraorbital $s$-wave spin-singlet paring\cite{PhysRevX.1.011009}, namely
\begin{eqnarray}\label{SM_iron2}
\mathcal{H}_{\text{sc}} = \Delta_0 + 2 \Delta_1 ( \cos k_x + \cos k_y ).
\end{eqnarray}
This form can generate both conventional  and sign-changed s waves between inner and outer pockets. We consider the sign-changed s wave to confirm its nontrivial topology.   Taking $\Delta_0 = 232$ meV and $\Delta_1 = 66.7$ meV,   the superconducting state is  the sign-changed s-wave pairing state with a minimum gap around $2.7$mev, as shown in Fig.\ref{fig_ironedge_SM}(a).

With the above parameters, we calculate the edge modes on the $(10)$ boundary shown in Fig.\ref{fig_ironedge_SM}(b).  There are  two degenerate Majorana cones, which is consistent with the analysis in the main text. Therefore, we can conclude that the sign-changed s-wave state in the iron-selenides is a second-order topological superconductor.


\begin{figure}[!htbp]
	\centering
	\includegraphics[width=0.7\linewidth]{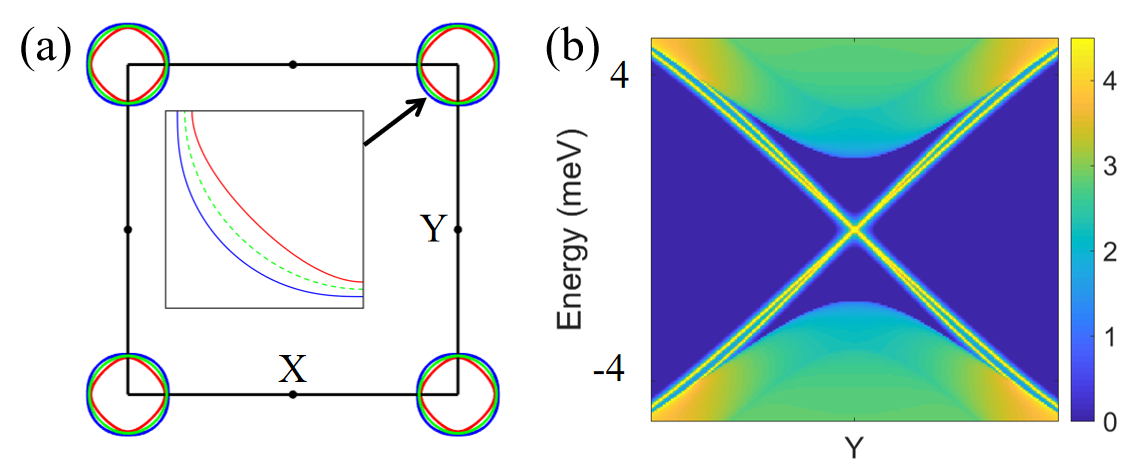}
	\caption{\label{fig_ironedge_SM} (color online) (a) shows the sign-changed s-wave pairing simulated by the superconducting order in Eq.\ref{SM_iron2}, with the sign of the superconducting order indicated by red ($-$) and blue ($+$). The Fermi surfaces are in the same condition with that in Fig.\ref{fig_iron_SM}(d), and the pairing nodes are label by the green color. (b) shows the edge modes on the $(10)$ boundary, corresponding to the superconducting state in (a).
	}
\end{figure}


\end{widetext}

\end{document}